\documentclass[reprint,prl, superscriptaddress,
nofootinbib]{revtex4-2}
\usepackage[normalem]{ulem}
\usepackage{amsmath}
\usepackage{amssymb}
\usepackage{graphicx}
\usepackage{dcolumn}
\usepackage{bm}

\usepackage{graphicx}
\usepackage{color}

\newcommand{\qq}{\begin{eqnarray}}
\newcommand{\qqq}{\end{eqnarray}}

\newcommand{\bfx}{\mathbf{x}}
\newcommand{\bfr}{\mathbf{r}}
\newcommand{\bff}{\mathbf{f}}

\newcommand{\bfv}{{\bf v}}

\newcommand{\bfq}{{\bf q}}

\newcommand{\AM}[1]{\textcolor{magenta}{#1}}

\begin{document}

\preprint{APS/123-QED}

\title{Interface dynamics of wet active systems}

\author{Fernando Caballero}
 \affiliation{Department of Physics, Brandeis University, Waltham, MA 02453, USA}
 \email{fcaballero@brandeis.edu}
\author{Ananyo Maitra}%
\affiliation{Laboratoire de Physique Th\'eorique et Mod\'elisation, CNRS UMR 8089, CY Cergy Paris Universit\'e, F-95032 Cergy-Pontoise Cedex, France}
\affiliation{Laboratoire Jean Perrin, UMR 8237 CNRS, Sorbonne Universit\'e, 75005 Paris, France}
\email{nyomaitra07@gmail.com}
\author{Cesare Nardini}
\affiliation{Service de Physique de l'Etat Condens\'e, CEA, CNRS Universit\'e Paris-Saclay, CEA-Saclay, 91191 Gif-sur-Yvette, France}
\affiliation{Sorbonne Universit\'e, CNRS, Laboratoire de Physique Th\'eorique de la Mati\`ere Condens\'ee, LPTMC, F-75005 Paris, France}
\email{cesare.nardini@gmail.com }

\begin{abstract}
We study the roughening of interfaces in phase-separated active suspensions on substrates. 
At both large length and timescales, we show that the interfacial dynamics belongs to the $|{\bf q}|$KPZ universality class discussed in Besse et al. Phys. Rev. Lett. {\bf 130}, 187102 (2023). This holds despite the presence of long-ranged fluid flows. At early times, however, or for sufficiently small systems, the roughening exponents are the same as those in the presence of a momentum-conserving fluid. Surprisingly, when the effect of substrate friction can be ignored, the interface becomes random beyond a de Gennes-Taupin lengthscale which depends on the interfacial tension.
\end{abstract}

\maketitle
The scale-free behavior, or roughening, of interfaces is a well-studied problem in statistical mechanics~\cite{krug1997origins,bray1994theory}. Early theoretical investigations~\cite{peters1979radius,plischke1984active,jullien1985scaling} centred on the Eden model~\cite{eden1958probabilistic}, originally proposed to describe the shape of cell colonies, and the ballistic deposition model~\cite{family1985scaling}.
The Kardar-Parisi-Zhang (KPZ) universality class \cite{forster1977large, kardar1986dynamic} was a major breakthrough, unifying the large scale behavior of a large set of different processes \cite{tauber2014critical}. 
However, a large class of interfaces---those with a conserved average height---cannot be described by the KPZ equation~\cite{krug1997origins, sun1989dynamics, caballero2018strong}.

Examples of this kind are naturally found in phase-separated systems. Already at equilibrium, since diffusive or fluid fluxes in the bulk relax much faster than the low wavenumber fluctuations of the interface itself, the interfacial dynamics becomes nonlocal and, even at linear level, does not obey Edwards-Wilkinson scaling~\cite{bray2001interface,shinozaki1993dispersion}. Specifically, a small amplitude height fluctuation $h_\bfq$ on wavevector $\bfq$ over a flat $d$-dimensional plane obeys
\qq\label{eq:old_linear}
\partial_th_\bfq(t) = -\gamma|\bfq|^ah_\bfq(t) + \sqrt{2D|\bfq|^{-2+a}}\xi_\bfq,
\qqq
where $\xi_\bfq$ is a Gaussian white noise and $\gamma$ is proportional to interfacial tension. When the underlying dynamics of the density field is diffusive, $a=3$; $a=1$ for systems dominated by hydrodynamic effects. Importantly, in both cases, the static structure factor of interface fluctuations depends on wavenumber as $S(\bfq)=\langle |h_\bfq|^2 \rangle\sim|\bfq|^{-2}$; this is a standard result of capillary wave theory~\cite{rowlinson2013molecular}. In equilibrium systems, Eq. \eqref{eq:old_linear} is believed to be exact for the large-scale properties in the steady state because there are no additional terms that are relevant in the renormalization group (RG) sense. Hence, the scale-free properties of the interface, commonly encoded in the roughening exponent $\chi$ and dynamical exponent $z$, that determine the long-range behavior of spatial and temporal correlations $\langle h(\bfx,t)h(\bfx',t)\rangle\sim |\bfx-\bfx'|^{2\chi}$ and $\langle h(\bfx,t)h(\bfx,t')\rangle\sim |t-t'|^{2\chi/z}$, are given by the linear theory ($z=a$, $\chi=(2-d)/2$). 
Analogous linear descriptions of the interface have been recently obtained for active systems in the dry~\cite{fausti2021capillary} and wet settings~\cite{caballero2022activity,gulati2024nonreciprocal}. However, in the presence of activity, a relevant non-linearity can singularly modify the dynamics of capillary waves and a new universality class, termed $|{\bf q}|$KPZ, emerges in the absence of any fluid flow~\cite{besse2023interface}.

A primary reason for developing active matter theories is to bring mechanobiology within the ambit of condensed matter physics, and the vast majority of microbiological systems require a fluid medium~\cite{marchetti2013hydrodynamics,julicher2018hydrodynamic}. Indeed, most experimentally studied active systems, of both biological and synthetic origins, are wet to some extent \cite{palacci2013living,sanchez2012spontaneous,bricard2013emergence,patteson2018propagation,adkins2022dynamics,tayar2023controlling, galajda2007wall}. In most of these, active particles move in quasi-two-dimensional geometries, and the fluid flow is screened; one might thus conclude that the dry limit applies. It is however known that, even in the presence of screening, the fluid flows generated by active forces are long-ranged, decaying algebraically as $1/r^3$~\cite{diamant2009hydrodynamic}, where $r$ is the distance between the point at which the force is applied and the one where the flow is observed. Hence, even in the presence of substrate friction, and at sufficiently large scales, 
it is unclear whether the fully dry physics is recovered. This does not happen in other circumstances, such as for polar flocks: fluid flows, even when screened, have profound effects, for instance, eliminating~\cite{maitra2020swimmer, CLT_Ncom, Bricard, Brotto} or modifying~\cite{Sarkar_Toner} giant number fluctuations.

In this Letter, we examine the physics of interfaces in active, isotropic suspensions in the presence of fluid flows. Our setup applies to active suspensions either on a substrate or in confined geometries, thus encompassing most experimentally studied realisations of active systems. 
We demonstrate that an arbitrarily small activity qualitatively modifies the long-wavelength, low-frequency scaling of the 
spatiotemporal correlators of one or two-dimensional interfaces; i.e., activity is a relevant perturbation in the renormalization group sense. This is due to the presence of an activity-induced nonlinearity.
Because of this, 
we show that these interfaces belong to the $|\bfq|$KPZ universality class at sufficiently large scales and that the amplitude of the active force can be used as a control parameter to tune the strength of the $|{\bf q}|$KPZ non-linearity.
Intriguingly, in wavenumber regimes in which the fluid dynamics is not screened, the interface normals become uncorrelated over large scales, and the growth does not follow a power law in time unlike in standard roughening scenarios; in this regime, the stationary structure factor becomes $S(\bfq)\sim |\bfq|^{-3}$, also at odds with the standard capillary-wave theory. Finally, for the specific case in which momentum is exactly conserved and hence the suspension is globally force-free, we find that the interface dynamics remains linear for the class of systems we considered. 

We now show how we obtain these results. Our starting point is the natural description of the density field $\phi(\bfr,t)$ driven by advective and diffusive dynamics:
\qq\label{eq:modelh_base}
\partial_t\phi(\bfr,t) +\bfv\cdot\nabla\phi(\bfr,t) = M\nabla^2\mu + \sqrt{2MD}\nabla\cdot \mathbf{\Lambda}_n,
\qqq
where $M$ is the mobility, $\mathbf{\Lambda}_n$ is a zero-average uncorrelated noise vector with unit variance,
$\bfv$ is the incompressible velocity field ($\nabla\cdot{\bf v}=0$), and $\mu= \delta F/\delta\phi$ is the chemical potential associated with the free energy $F$. The latter is a Landau-Ginzburg expansion in the phase field and its gradients
$
F = \int d\bfr \,\Big[f(\phi) + (k/2)(\nabla\phi)^2\,\Big],
$
where the bulk free energy density $f(\phi)$ has a double well structure, and where the gradient term represents energetic penalties to density gradients.
The flow field $\bfv$ obeys 
\qq \label{eq:momentumeq}
0= \eta\nabla^2\bfv -\Gamma\bfv -\nabla P + \bff + \bff^n,
\qqq
in the low-Reynolds number regime;
here $\eta$ is the viscosity, $P$ is a pressure determined by incompressibility, $\bff$ is the total deterministic force density arising from density modulations,  and $\Gamma$ represents frictional dampening of the flow. The noise term $\bff^n$ has zero average and a variance which, in Fourier space, reads 
\qq
&\langle f^n_i({\bf p},\omega)f^n_j({\bf p}',\omega')\rangle = \\  &(2\pi)^{d+2}2D(\Gamma+\eta|{\bf p}|^2)
\delta_{ij}
\delta({\bf p}+{\bf p}')\delta(\omega+\omega')\,,\nonumber
\qqq
where ${\bf p}=(\bfq,q_y)$ is the wave-vector in $d+1$ dimensional space and $i$ denotes the spatial components.
For simplicity, we assume that $\bff^n$ obeys a fluctuation-dissipation relation even in our active system.

Upon setting $\bff =\bff^p = -\phi\nabla\mu$ and $\Gamma=0$, Eqs. \eqref{eq:modelh_base} and \eqref{eq:momentumeq} reduce to model H, the standard description of phase-separation in momentum-conserved systems~\cite{cates2018theories, hohenberg1977theory}. 
Notice that $\Gamma\neq0$ implies that momentum is not conserved. This naturally arises in two cases: either due to friction of the active fluid with the substrate or, even when this is strictly absent, if the fluid is confined vertically by no-slip walls~\cite{diamant2009hydrodynamic, Oron}. In the latter case, once the vertical motion is averaged out, the ensuing effective two-dimensional description leads to Eq. (\ref{eq:momentumeq}), and the screening length $1/\lambda=\sqrt{\eta/\Gamma}$ turns out to be proportional to the film thickness~\cite{diamant2009hydrodynamic, AM_nem, maitra2020swimmer, Brotto, Oron}. 
With a slight abuse of terminology, we will keep referring to $\Gamma$ as the friction in either case. It is crucial for the following discussion that, despite screening, the fluid flow disturbances still decay algebraically for $r\gg1/\lambda$: friction only renormalises the decay from a $1/r^2$ scaling in unbounded suspensions to a $1/r^3$ in the present case~\cite{diamant2009hydrodynamic}, implying that the suspension still interacts non-locally. 


We now describe how to introduce activity in Eqs. (\ref{eq:modelh_base}) and (\ref{eq:momentumeq}). As in previous attempts at describing wet active suspensions~\cite{tiribocchi2015active,cates2018theories}, activity breaks the relation between $\bf f$ and the gradients of the chemical potential so that ${\bf f}= {\bf f}^p+{\bf f}^a$. If momentum is conserved, the force density must come from a stress tensor $\sigma_{ij}$, with $f^a_i=\nabla_j\sigma_{ij}$. Expanding the stress tensor to leading order in gradients yields $\sigma_{ij}= \kappa (\nabla_i\phi\nabla_j\phi-\delta_{ij}|\nabla\phi|^2/(d+1))$, a model known as the Active Model H~\cite{tiribocchi2015active}. When $\Gamma\neq0$, active forces that cannot be written as the divergence of a local stress tensor are also present. In this case, including all terms up to order $\mathcal{O}(\nabla^3\phi^3)$ in $\bff^a$, we find
\qq\label{eq:f_i-active}
f^a_i=  \AM{\nabla_j\sigma_{ij}}+
K_1\phi(\nabla^2\phi)\nabla_i\phi+ K_2 |\nabla\phi|^2\nabla_i \phi\,.
\qqq

A distinct way to introduce activity is to use a force density that pushes the interface in the direction normal to it (this is particularly natural if the interface is a membrane hosting active inclusions~\cite{prost1996shape,cagnetta2022universal,ramaswamy2000nonequilibrium}). In writing this force density, we also introduce the coordinate system that we use throughout the rest of this work. We will assume without loss of generality that the interface is normal to the $y$ direction, and denote by $\bfx$ the directions parallel to it. As a function of these coordinates, we can write the active force density as
\qq\label{eq:active_force}
{\bf f}^a=2\alpha_f\hat{{\bf n}}\delta(y-h),
\qqq
where $\hat{{\bf n}} = (\hat{y}-\nabla h)/\sqrt{1+(\nabla h)^2}$ is the local interface normal. This force density is such that its strength is proportional to $\alpha_f$, and points from the dense to the dilute regions~\cite{footnote}. We will show that introducing activity either via Eq. (\ref{eq:f_i-active}) or Eq. (\ref{eq:active_force}) leads to the same results, thereby demonstrating that our conclusions are generically valid.

In the case of model H, the interface height is described by Eq. \eqref{eq:old_linear} with $a=1$~\cite{bray2001interface,shinozaki1993dispersion}. This can be shown by neglecting diffusive fluxes (i.e. setting $M=0$ in Eq. (\ref{eq:modelh_base})), which contribute at higher order in gradients, and assuming that the density profile $\varphi$ across the interface evolves quasistatically with respect to fluctuations of the interfacial height. To leading order in $\bf q$ and $h$, and in the absence of overhangs, we can assume the interfacial profile to satisfy
\qq\label{eq:Ansatz}
\phi({\bf x},y,t)=\varphi(y-h({\bf x},t))\,,
\qqq
where $\varphi$ is the density profile across a flat interface. 
We now use the same technique to investigate the effect of the active forces and friction. 
We formally solve 
Eq. \eqref{eq:modelh_base} with $M=0$ to yield the following $\phi$ dynamics in closed form:
\qq\label{eq:phi_formal}
\partial_t\phi(\bfr,t) = - \partial_i\phi(\bfr,t)\int d\bfr' T_{ij}(\bfr-\bfr')f_j(\bfr') + \tilde{\xi}(\bfr),
\qqq
where the Gaussian noise $\tilde{\xi}$ has zero average and the variance
\qq\label{eq:noise-interface}
\langle\tilde{\xi}(\bfr,t)\tilde{\xi}(\bfr',t)\rangle = 2D\partial_i\phi(\bfr)T_{ij}(\bfr-\bfr')\partial_j\phi(r')\delta(t-t').
\qqq
Here, $T_{ij}$ are the components of a modified Oseen tensor, which are most conveniently written in Fourier space as 
\qq
{\tilde T}_{ij}({\bf p}) = \frac{1}{\Gamma + \eta|{\bf p}|^2}\left(\delta_{ij} - \frac{{\bf p}_i{\bf p}_j}{|{\bf p}|^2}\right).
\qqq

 \begin{figure}
    \centering
    \includegraphics[width=0.4\textwidth]{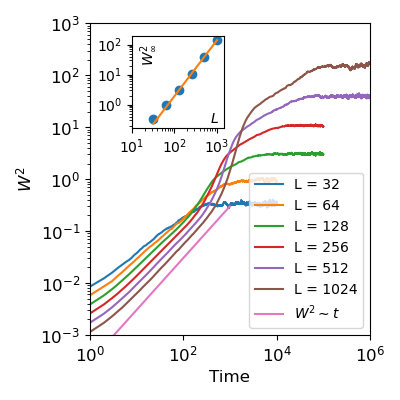}
    \caption{Interface width $W^2$ as a function of time as obtained from numerical integration of Eq. \eqref{eq:main_result} for screening length $1/\lambda = 1$ and $\bar{\alpha}=0.63$. The inset shows the saturation values $W_\infty^2$ as a function of system size $L$, with a power law fit with slope $1.85$.}
    \label{fig:lambda1}
\end{figure}

\begin{figure*}
\includegraphics[width=\textwidth]{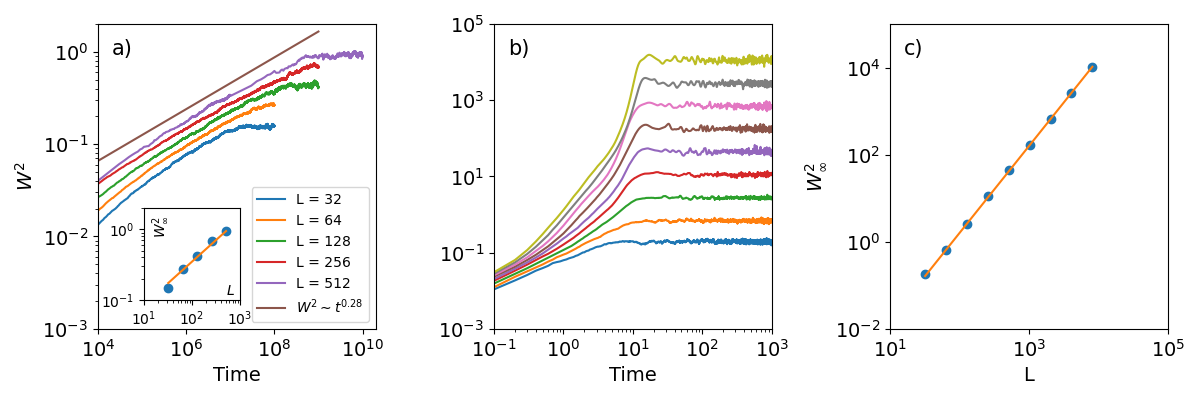}
\caption{Evolution of the width in two limits; a) In the fully screened limit, with $\bar\alpha=0.6$,  $\lambda = 10^3$ and sufficiently large system-size, we observe a growth law compatible with the $|\bfq|$KPZ universality class ($W^2\sim t^{0.28}$). The inset shows the saturation values $W_\infty^2$ as a function of system size $L$, and a fit to a power law gives a slope of $0.61\pm0.04$, compatible with the $|\bfq|$KPZ universality class. b) In the unscreened limit, with $\lambda=0$ (other parameters as in panel a), each line from bottom to top corresponds to systems sizes from $L=2^5$ to $L=2^{13}$ in powers of $2$.), the width displays an initially linear growth as in fully wet systems before the effect of the nonlinearity triggers a growth faster than a power law. The saturation time does not seem to depend on system size, at odds with the standard roughening scenario. c) Values of the saturated width as a function of system size corresponding to the unscreened limit in perfect agreement with our analytical prediction that $\chi=1$ (orange line).}
\label{fig:widths}
\end{figure*}

The only component of the Oseen tensor relevant to our treatment below is $T_{yy}$: the contributions to Eq. (\ref{eq:phi_formal}) from all other components either vanish, turn out to be irrelevant in the RG sense, or can be reabsorbed in pressure jumps across the interface. It is useful to write $T_{yy}$ in real space in $y$ and Fourier space in $\bfx$, which we denote as $\mathcal T_{yy}(\bfq,y)$:
\qq\label{eq:Oseen-q-y}
\mathcal T_{yy}(\bfq,y) = & \frac{|\bfq| e^{-|\bfq||y|}}{2 \eta \lambda^2}-\frac{|\bfq|^2 e^{-|y| \sqrt{\lambda ^2+|\bfq|^2}}}{2 \eta \lambda^2 \sqrt{\lambda ^2+|\bfq|^2}}.
\qqq
Substituting Eq. \eqref{eq:active_force} in Eq. \eqref{eq:phi_formal}, and expanding to lowest nonlinear order in $h$, we obtain the main analytical result of this Letter~\cite{SM}:
\qq\label{eq:main_result}
\partial_t h_\bfq &= 
-\gamma|\bfq|^2 \mathcal T_{yy}(\bfq,0)h_\bfq +\alpha T_{yy}(\bfq,0) \mathcal{F}
[|\nabla h|^2]
+\xi_\bfq\,,
\qqq
where $\mathcal{F}$ is the Fourier transform operator, and the Gaussian noise has zero average and correlations
\qq\label{eq:noise-effective-interface-eq}
\langle\xi_\bfq(t)\xi_{\bfq'}(t')\rangle = 2D(2\pi)^d\mathcal T_{yy}(\bfq,0)\delta_{\bfq,-\bfq'}\delta(t-t')\,.
\qqq
The nonlinear term proportional to $\alpha$ closely resembles the KPZ non-linearity but, crucially, comes with a ${\bf q}$-dependent prefactor encoded in the Oseen tensor. 
When the active force density is given by Eq. \eqref{eq:active_force}, we find $\alpha=\alpha_f$~\cite{SM}. When the activity given by Eq. \eqref{eq:f_i-active}, we show that $\alpha = (K_1-2K_2)\int du \varphi(u)\varphi'(u)\varphi''(u)$, where the integral is taken across the interface. In both cases, the amplitude of the leading non-linear term in Eq. (\ref{eq:main_result}) is directly proportional to the strength of the active force densities acting on the fluid. In Eq. (\ref{eq:main_result}) we have neglected nonlinearities that are higher order in $h$ and are thus irrelevant in the RG sense. 

We first investigate Eq.~\eqref{eq:main_result} numerically for one-dimensional interfaces ($d=1$) using the package recently developed in~\cite{caballero2024cupss}. This is a pseudo-spectral scheme with a computational complexity proportional to $N\log N$, where $N$ is the number of collocation points, despite the non-local nature of Eq. (\ref{eq:main_result}). We monitor the interface width $W^2 = L^{-1}\int_0^L dx' (h-\bar h)^2$, where $\bar h$ is the average of $h(x,t)$ across the system. We then measure the scaling exponents $\chi$ and $z$ using the fact that $W^2\sim t^{2\chi/z}$ and that, once the interface width has saturated in time to its stationary value, $W^2_\infty$, $W^2_\infty\sim L^{2\chi}$. These measures are standard in studies of roughening phenomena.

We integrate Eq. \eqref{eq:main_result} for several system sizes, and average the results over multiple noise realizations (from $10^2$ to $10^3$ depending on system size). When $\alpha=0$, our numerical measures yield the growth predicted by dimensional analysis, which corresponds to $z=1$ and $\chi=1/2$~\cite{SM}. We then 
performed simulations with  $\bar{\alpha} \sim \mathcal{O}(1)$, where $\bar{\alpha}=\alpha \sqrt{D/\gamma^3}$ is the non-dimensional coupling setting the strength of the nonlinearity. Fig. 
\ref{fig:lambda1} shows the width as a function of time and various system sizes for $\bar{\alpha}=0.63$ and screening length $1/\lambda=1$. 
At early times and large enough system sizes, we find $z\simeq 1$, which is the expected value for momentum conserving fluids~\cite{bray1994theory,shinozaki1993dispersion}; a linear theory suffices to obtain this result. The nonlinearity then kicks in and creates a regime of rapid growth, showing a clear deviation from classical roughening. 
For large system sizes ($L\geq 512$) and late times, $W^2$ seems to enter a second power-law regime before saturating, but 
we were unable to obtain a clear measure of the associated exponent up to system sizes that are computationally accessible. Finally, the width saturates to a value that depends on $L$, as expected in a standard roughening scenario. The associated measure for $\chi$ is shown in the inset of Fig. \ref{fig:lambda1} and is compatible with $2\chi\simeq 1.85$, greater than the value expected from the linear theory ($2\chi=1$). We however show below that this regime is only a transient effect due to finite system-size.

An accurate numerical study of Eq. \eqref{eq:main_result} showing the scaling at both early and late times for a single set of parameters
is computationally prohibitive. Therefore, to gain insight into the various regimes, we consider different limits of Eq. \eqref{eq:main_result} and investigate them both analytically and numerically. First, we show that friction is dominant at late enough times and small enough wavenumbers $|{\bf q}|$. Indeed, for $|{\bf q}|\ll \lambda$, one can show that 
\qq
\mathcal T_{yy}(\bfq,0)\to_{|{\bf q}\ll \lambda} \frac{|\bfq|}{2\eta \lambda^2}\,
\qqq
and hence Eq. \eqref{eq:main_result} becomes
\qq
\label{eq:qKPZ}
\partial_th_\bfq(t) = - \frac{\gamma|\bfq|^3}{2\Gamma}h_\bfq(t)+\frac{\alpha|\bfq|}{2\Gamma}\mathcal F[(\nabla h(x,t))^2]+\xi_\bfq\label{eq:qkpc}\,.
\qqq
Eq. (\ref{eq:qKPZ}) is the $|\bfq|$KPZ equation, first introduced in \cite{besse2023interface}.
This result is remarkable and merits further comments. First, while it is expected that capillary waves relax on a timescale proportional to $1/|\bfq|^3$ when the bulk dynamics of the mass density is diffusive~\cite{bray1994theory,fausti2021capillary}, it is surprising to find the same relaxation timescale in our case, in which the bulk dynamics is dominated by long-ranged fluid flows.
Second, our wet setting allows us to connect the active forces directly to the amplitude of the nonlinearity in the interface equation; this is at odds with systems that have diffusive bulk dynamics, in which the $|\bfq|$KPZ nonlinearity only arises as an effect of fluctuations along the RG flow~\cite{besse2023interface}. 

To confirm our prediction that the interface belongs to the $|{\bf q}|$KPZ universality class in the presence of friction, we integrated Eq. \eqref{eq:main_result} numerically for strong friction. The evolution of the interface width $W^2$ with time is reported in Fig. \ref{fig:widths}(a) for $\lambda=10^3$: While in the linear case, we observe a mean-field growth $W^2\sim t^{1/3}$~\cite{SM}, for sufficiently large systems ($L\gtrsim 256$) and late times the growth is consistent with $2\chi/z\approx 0.28$, the value predicted for the $|\bfq|$KPZ universality class. 

Having described the large-scale behaviour of Eq. \eqref{eq:main_result}, we now rationalise the early-time regime. To do so, we consider the regime $\lambda\rightarrow0$ while retaining a finite $\alpha$, which we refer to as the unscreened limit in the following. From Eq. (\ref{eq:Oseen-q-y}), we obtain that 
\qq
\mathcal T_{yy}(\bfq,0)\to_{\lambda\to0} \frac{1}{4\eta |\bfq|}\,,
\qqq
and hence
 \eqref{eq:main_result} becomes
\qq
\partial_th_\bfq(t) = - \frac{\gamma|\bfq|}{4\eta}h_\bfq(t)+\frac{\alpha}{4\eta|\bfq|}\mathcal F[(\nabla h(\bfx,t))^2]+\xi_\bfq\,.\label{eq:qm1kpz}
\qqq
The linear roughening properties of Eq. \eqref{eq:qm1kpz} can be read off by setting $\alpha = 0$, which gives $z=1$ and $2\chi=2-d$. Going beyond the linear regime, we show by a one-loop RG analysis that the non-linearity proportional to $\alpha$ generates a noise with a lower order correlation than the one in Eq. \eqref{eq:qm1kpz}. That is, the nonlinearity $\alpha$ generates $\xi'_\bfq(t)$ which has zero average and correlations
\qq\label{eq:unscreened-n}
\langle\xi'_\bfq(t)\xi'_{\bfq'}(t')\rangle = \frac{D'}{|\bfq|^2}(2\pi)^d\delta(\bfq+\bfq')\delta(t-t'),
\qqq
where $D'$ is a function of the dimensionless nonlinearity $\bar\alpha$ and of the ultraviolet cut-off $\Lambda$, see~\cite{SM} for the explicit expression. 

Eq. \eqref{eq:unscreened-n} implies a static structure factor $S(\bfq)\sim |\bfq|^{-3}$, and $\chi = (3-d)/2$. When $d=1$, this implies $\chi=1$, which is believed on general grounds to be the upper bound for the roughness exponent associated with interfaces without overhangs \cite{tauber2014critical}. 
Furthermore, the correlator $\langle |\delta{\bf n}(\bfx)-\delta {\bf n}({\bf 0})|^2\rangle$ for fluctuations of the normal $\delta{\bf n}\approx\nabla h$ now diverges logarithmically with distance, i.e. the interface becomes random beyond a de Gennes-Taupin lengthscale $\propto e^{\gamma/D'}$. Correspondingly, the normal, averaged over space, vanishes, $\lim_{L\to\infty} (1/L)\int d{\bf x}\, \delta{\bf n} =0$. Interestingly, while the original de Gennes-Taupin lengthscale was obtained for a tension-free membrane and depends on the bending rigidity~\cite{Taupin}, here it depends on the surface tension. We finally observe that $\chi=1$ in $d=1$ implies that the scaling dimension of $\nabla h$ vanishes, implying that any power of $\nabla h$ becomes equally relevant in the RG sense. 


To verify our predictions for the early time dynamics of the interfaces, we studied Eq. (\ref{eq:qm1kpz}) numerically in the unscreened limit. Fig. \ref{fig:widths}b reports the time evolution of the interfacial width. At early times, $W^2\sim t$, which corresponds to the linear behaviour of Eq. \eqref{eq:main_result}, before that the non-linearity induces a faster growth which does not seem to follow a power law. This is compatible with what was already observed by numerically integrating the full interface equation (\ref{eq:main_result}), see Fig. \ref{fig:lambda1}, confirming that the unscreened limit describes the early-time dynamics of the interface. Furthermore, the scaling of the saturated width with $L$ fits a power law $W^2\sim L^{1.99}$, extremely close to the prediction previously obtained analytically ($\chi=1$, corresponding to $W_\infty^2\sim L^2$). 
Interestingly, the timescale at which $W^2$ saturates is found to be system-size independent, a fact that is not explained by the standard roughening scenario.

We finally consider the regime in which the system is fully wet. In this case, in addition to $\lambda$ being $0$, all force densities entering the Stokes equation arise from the divergence of a stress tensor. \AM{This rules out the active forces with coefficients $K_1$ and $K_2$} in Eq. (\ref{eq:f_i-active}) except when $K_1-2K_2=0$ (in which case the associated stress tensor is $\sigma_{ij}^a=\phi(\nabla_i\phi\nabla_j\phi-\delta_{ij}(\nabla\phi)^2/2)$). 
More generally, we consider the most general stress tensor to order $\mathcal{O}(\nabla^2)$ and any order in the density: $\sigma_{ij}=-\kappa(\phi)\partial_i\phi\partial_j\phi$, with $\kappa(\phi)$ being an arbitrary function of $\phi$. This includes Active Model H as a special case when $\kappa$ is constant and the active forces in Eq. (\ref{eq:f_i-active}) when $K_1-2K_2=0$. Following the same steps as those for obtaining Eq. (\ref{eq:main_result}), we show that $\alpha=0$ in this case~\cite{SM}: with full momentum conservation, we only find non-linearities that are irrelevant from an RG point of view. 
Note however that the non-linearity proportional to $\alpha$ might be produced by even more general forms of $\sigma_{ij}$, or by fluctuations along the RG flow once irrelevant non-linearities are included. If not, the interface equation in the fully wet limit is linear even in an active setting and is given by eq. (\ref{eq:old_linear}) with $a=1$.

In conclusion, we have described interfaces in wet phase-separated active systems. In the presence of fluid friction, we have shown that they belong, at sufficiently large scales, to the $|{\bf q}|$KPZ universality class. This result is surprising because the fluid flow, although screened, induces long-range interactions that are non-diffusive. We have further shown that the early-time dynamics of these interfaces are analogous to those in momentum-conserved systems. The two regimes are connected by a fast growth that is not described by the standard roughening scenario but can be rationalised by neglecting the friction between the substrate and the fluid; in this regime, the static structure factor is not described by the standard capillary-wave theory. 

It would be interesting to investigate the relevance of our work for confluent biological tissues: although we expect the same fundamental physics as the one described here, interfaces are significantly sharper than in fluids~\cite{sussman1,sussman2}.
Our results may also prove relevant for recent experiments on phase-separated active fluids, which are currently able to probe scales at which the system is isotropic on average~\cite{adkins2022dynamics,zhao2024asymmetric}. Techniques similar to those employed here may be used to describe shape fluctuations of biomolecular condensates~\cite{law2023bending,caragine2018surface}, in which the effect of fluid flow has only recently been investigated~\cite{seyboldt2018role,galvanetto2023extreme}.


\begin{acknowledgments}
FC thanks Cristina Marchetti, Paarth Gulati and Aparna Baskaran for useful discussions. FC acknowledges support from the NSF, grants DMR-2041459 \& DMR-2011846. AM and CN  acknowledge the
support of the ANR grant PSAM. CN acknowledges the support of the INP-IRP grant IFAM and from the Simons Foundation. AM and CN thank the Isaac Newton Institute for Mathematical Sciences for support and hospitality during the program ``Anti-diffusive dynamics: from sub-cellular to astrophysical scales'' when part of this work was undertaken. This research was supported in part by grant NSF PHY-2309135 to the Kavli Institute for Theoretical Physics (KITP). AM acknowledges a TALENT fellowship awarded by CY Cergy Paris Universit\'e.
\end{acknowledgments}
\bibliographystyle{apsrev4-1}
\bibliography{apssamp}

\begin{thebibliography}{53}%
\makeatletter
\providecommand \@ifxundefined [1]{%
 \@ifx{#1\undefined}
}%
\providecommand \@ifnum [1]{%
 \ifnum #1\expandafter \@firstoftwo
 \else \expandafter \@secondoftwo
 \fi
}%
\providecommand \@ifx [1]{%
 \ifx #1\expandafter \@firstoftwo
 \else \expandafter \@secondoftwo
 \fi
}%
\providecommand \natexlab [1]{#1}%
\providecommand \enquote  [1]{``#1''}%
\providecommand \bibnamefont  [1]{#1}%
\providecommand \bibfnamefont [1]{#1}%
\providecommand \citenamefont [1]{#1}%
\providecommand \href@noop [0]{\@secondoftwo}%
\providecommand \href [0]{\begingroup \@sanitize@url \@href}%
\providecommand \@href[1]{\@@startlink{#1}\@@href}%
\providecommand \@@href[1]{\endgroup#1\@@endlink}%
\providecommand \@sanitize@url [0]{\catcode `\\12\catcode `\$12\catcode `\&12\catcode `\#12\catcode `\^12\catcode `\_12\catcode `\%12\relax}%
\providecommand \@@startlink[1]{}%
\providecommand \@@endlink[0]{}%
\providecommand \url  [0]{\begingroup\@sanitize@url \@url }%
\providecommand \@url [1]{\endgroup\@href {#1}{\urlprefix }}%
\providecommand \urlprefix  [0]{URL }%
\providecommand \Eprint [0]{\href }%
\providecommand \doibase [0]{http://dx.doi.org/}%
\providecommand \selectlanguage [0]{\@gobble}%
\providecommand \bibinfo  [0]{\@secondoftwo}%
\providecommand \bibfield  [0]{\@secondoftwo}%
\providecommand \translation [1]{[#1]}%
\providecommand \BibitemOpen [0]{}%
\providecommand \bibitemStop [0]{}%
\providecommand \bibitemNoStop [0]{.\EOS\space}%
\providecommand \EOS [0]{\spacefactor3000\relax}%
\providecommand \BibitemShut  [1]{\csname bibitem#1\endcsname}%
\let\auto@bib@innerbib\@empty
\bibitem [{\citenamefont {Krug}(1997)}]{krug1997origins}%
  \BibitemOpen
  \bibfield  {author} {\bibinfo {author} {\bibfnamefont {J.}~\bibnamefont {Krug}},\ }\href@noop {} {\bibfield  {journal} {\bibinfo  {journal} {Adv. Phys.}\ }\textbf {\bibinfo {volume} {46}},\ \bibinfo {pages} {139} (\bibinfo {year} {1997})}\BibitemShut {NoStop}%
\bibitem [{\citenamefont {Bray}(1994)}]{bray1994theory}%
  \BibitemOpen
  \bibfield  {author} {\bibinfo {author} {\bibfnamefont {A.~J.}\ \bibnamefont {Bray}},\ }\href@noop {} {\bibfield  {journal} {\bibinfo  {journal} {Adv. Phys.}\ }\textbf {\bibinfo {volume} {43}},\ \bibinfo {pages} {357} (\bibinfo {year} {1994})}\BibitemShut {NoStop}%
\bibitem [{\citenamefont {Peters}\ \emph {et~al.}(1979)\citenamefont {Peters}, \citenamefont {Stauffer}, \citenamefont {H{\"o}lters},\ and\ \citenamefont {Loewenich}}]{peters1979radius}%
  \BibitemOpen
  \bibfield  {author} {\bibinfo {author} {\bibfnamefont {H.}~\bibnamefont {Peters}}, \bibinfo {author} {\bibfnamefont {D.}~\bibnamefont {Stauffer}}, \bibinfo {author} {\bibfnamefont {H.}~\bibnamefont {H{\"o}lters}}, \ and\ \bibinfo {author} {\bibfnamefont {K.}~\bibnamefont {Loewenich}},\ }\href@noop {} {\bibfield  {journal} {\bibinfo  {journal} {Z. Phys. B Con. Mat.}\ }\textbf {\bibinfo {volume} {34}},\ \bibinfo {pages} {399} (\bibinfo {year} {1979})}\BibitemShut {NoStop}%
\bibitem [{\citenamefont {Plischke}\ and\ \citenamefont {R{\'a}cz}(1984)}]{plischke1984active}%
  \BibitemOpen
  \bibfield  {author} {\bibinfo {author} {\bibfnamefont {M.}~\bibnamefont {Plischke}}\ and\ \bibinfo {author} {\bibfnamefont {Z.}~\bibnamefont {R{\'a}cz}},\ }\href@noop {} {\bibfield  {journal} {\bibinfo  {journal} {Phys. Rev. Lett.}\ }\textbf {\bibinfo {volume} {53}},\ \bibinfo {pages} {415} (\bibinfo {year} {1984})}\BibitemShut {NoStop}%
\bibitem [{\citenamefont {Jullien}\ and\ \citenamefont {Botet}(1985)}]{jullien1985scaling}%
  \BibitemOpen
  \bibfield  {author} {\bibinfo {author} {\bibfnamefont {R.}~\bibnamefont {Jullien}}\ and\ \bibinfo {author} {\bibfnamefont {R.}~\bibnamefont {Botet}},\ }\href@noop {} {\bibfield  {journal} {\bibinfo  {journal} {J. Phys. A-Math. Gen.}\ }\textbf {\bibinfo {volume} {18}},\ \bibinfo {pages} {2279} (\bibinfo {year} {1985})}\BibitemShut {NoStop}%
\bibitem [{\citenamefont {Eden}(1958)}]{eden1958probabilistic}%
  \BibitemOpen
  \bibfield  {author} {\bibinfo {author} {\bibfnamefont {M.}~\bibnamefont {Eden}},\ }in\ \href@noop {} {\emph {\bibinfo {booktitle} {Symposium on information theory in biology}}}\ (\bibinfo {organization} {Pergamon Press, New York},\ \bibinfo {year} {1958})\ pp.\ \bibinfo {pages} {359--370}\BibitemShut {NoStop}%
\bibitem [{\citenamefont {Family}\ and\ \citenamefont {Vicsek}(1985)}]{family1985scaling}%
  \BibitemOpen
  \bibfield  {author} {\bibinfo {author} {\bibfnamefont {F.}~\bibnamefont {Family}}\ and\ \bibinfo {author} {\bibfnamefont {T.}~\bibnamefont {Vicsek}},\ }\href@noop {} {\bibfield  {journal} {\bibinfo  {journal} {J. Phys. A-Math. Gen.}\ }\textbf {\bibinfo {volume} {18}},\ \bibinfo {pages} {L75} (\bibinfo {year} {1985})}\BibitemShut {NoStop}%
\bibitem [{\citenamefont {Forster}\ \emph {et~al.}(1977)\citenamefont {Forster}, \citenamefont {Nelson},\ and\ \citenamefont {Stephen}}]{forster1977large}%
  \BibitemOpen
  \bibfield  {author} {\bibinfo {author} {\bibfnamefont {D.}~\bibnamefont {Forster}}, \bibinfo {author} {\bibfnamefont {D.~R.}\ \bibnamefont {Nelson}}, \ and\ \bibinfo {author} {\bibfnamefont {M.~J.}\ \bibnamefont {Stephen}},\ }\href@noop {} {\bibfield  {journal} {\bibinfo  {journal} {Phys. Rev. A}\ }\textbf {\bibinfo {volume} {16}},\ \bibinfo {pages} {732} (\bibinfo {year} {1977})}\BibitemShut {NoStop}%
\bibitem [{\citenamefont {Kardar}\ \emph {et~al.}(1986)\citenamefont {Kardar}, \citenamefont {Parisi},\ and\ \citenamefont {Zhang}}]{kardar1986dynamic}%
  \BibitemOpen
  \bibfield  {author} {\bibinfo {author} {\bibfnamefont {M.}~\bibnamefont {Kardar}}, \bibinfo {author} {\bibfnamefont {G.}~\bibnamefont {Parisi}}, \ and\ \bibinfo {author} {\bibfnamefont {Y.-C.}\ \bibnamefont {Zhang}},\ }\href@noop {} {\bibfield  {journal} {\bibinfo  {journal} {Phys. Rev. Lett.}\ }\textbf {\bibinfo {volume} {56}},\ \bibinfo {pages} {889} (\bibinfo {year} {1986})}\BibitemShut {NoStop}%
\bibitem [{\citenamefont {T{\"a}uber}(2014)}]{tauber2014critical}%
  \BibitemOpen
  \bibfield  {author} {\bibinfo {author} {\bibfnamefont {U.~C.}\ \bibnamefont {T{\"a}uber}},\ }\href@noop {} {\emph {\bibinfo {title} {Critical dynamics: a field theory approach to equilibrium and non-equilibrium scaling behavior}}}\ (\bibinfo  {publisher} {Cambridge University Press},\ \bibinfo {year} {2014})\BibitemShut {NoStop}%
\bibitem [{\citenamefont {Sun}\ \emph {et~al.}(1989)\citenamefont {Sun}, \citenamefont {Guo},\ and\ \citenamefont {Grant}}]{sun1989dynamics}%
  \BibitemOpen
  \bibfield  {author} {\bibinfo {author} {\bibfnamefont {T.}~\bibnamefont {Sun}}, \bibinfo {author} {\bibfnamefont {H.}~\bibnamefont {Guo}}, \ and\ \bibinfo {author} {\bibfnamefont {M.}~\bibnamefont {Grant}},\ }\href@noop {} {\bibfield  {journal} {\bibinfo  {journal} {Phys. Rev. A}\ }\textbf {\bibinfo {volume} {40}},\ \bibinfo {pages} {6763} (\bibinfo {year} {1989})}\BibitemShut {NoStop}%
\bibitem [{\citenamefont {Caballero}\ \emph {et~al.}(2018)\citenamefont {Caballero}, \citenamefont {Nardini}, \citenamefont {van Wijland},\ and\ \citenamefont {Cates}}]{caballero2018strong}%
  \BibitemOpen
  \bibfield  {author} {\bibinfo {author} {\bibfnamefont {F.}~\bibnamefont {Caballero}}, \bibinfo {author} {\bibfnamefont {C.}~\bibnamefont {Nardini}}, \bibinfo {author} {\bibfnamefont {F.}~\bibnamefont {van Wijland}}, \ and\ \bibinfo {author} {\bibfnamefont {M.~E.}\ \bibnamefont {Cates}},\ }\href@noop {} {\bibfield  {journal} {\bibinfo  {journal} {Phys. Rev. Lett.}\ }\textbf {\bibinfo {volume} {121}},\ \bibinfo {pages} {020601} (\bibinfo {year} {2018})}\BibitemShut {NoStop}%
\bibitem [{\citenamefont {Bray}\ \emph {et~al.}(2001)\citenamefont {Bray}, \citenamefont {Cavagna},\ and\ \citenamefont {Travasso}}]{bray2001interface}%
  \BibitemOpen
  \bibfield  {author} {\bibinfo {author} {\bibfnamefont {A.~J.}\ \bibnamefont {Bray}}, \bibinfo {author} {\bibfnamefont {A.}~\bibnamefont {Cavagna}}, \ and\ \bibinfo {author} {\bibfnamefont {R.~D.}\ \bibnamefont {Travasso}},\ }\href@noop {} {\bibfield  {journal} {\bibinfo  {journal} {Phys. Rev. E}\ }\textbf {\bibinfo {volume} {65}},\ \bibinfo {pages} {016104} (\bibinfo {year} {2001})}\BibitemShut {NoStop}%
\bibitem [{\citenamefont {Shinozaki}(1993)}]{shinozaki1993dispersion}%
  \BibitemOpen
  \bibfield  {author} {\bibinfo {author} {\bibfnamefont {A.}~\bibnamefont {Shinozaki}},\ }\href@noop {} {\bibfield  {journal} {\bibinfo  {journal} {Phys. Rev. E}\ }\textbf {\bibinfo {volume} {48}},\ \bibinfo {pages} {1984} (\bibinfo {year} {1993})}\BibitemShut {NoStop}%
\bibitem [{\citenamefont {Rowlinson}\ and\ \citenamefont {Widom}(2013)}]{rowlinson2013molecular}%
  \BibitemOpen
  \bibfield  {author} {\bibinfo {author} {\bibfnamefont {J.~S.}\ \bibnamefont {Rowlinson}}\ and\ \bibinfo {author} {\bibfnamefont {B.}~\bibnamefont {Widom}},\ }\href@noop {} {\emph {\bibinfo {title} {Molecular theory of capillarity}}}\ (\bibinfo  {publisher} {Courier Corporation},\ \bibinfo {year} {2013})\BibitemShut {NoStop}%
\bibitem [{\citenamefont {Fausti}\ \emph {et~al.}(2021)\citenamefont {Fausti}, \citenamefont {Tjhung}, \citenamefont {Cates},\ and\ \citenamefont {Nardini}}]{fausti2021capillary}%
  \BibitemOpen
  \bibfield  {author} {\bibinfo {author} {\bibfnamefont {G.}~\bibnamefont {Fausti}}, \bibinfo {author} {\bibfnamefont {E.}~\bibnamefont {Tjhung}}, \bibinfo {author} {\bibfnamefont {M.}~\bibnamefont {Cates}}, \ and\ \bibinfo {author} {\bibfnamefont {C.}~\bibnamefont {Nardini}},\ }\href@noop {} {\bibfield  {journal} {\bibinfo  {journal} {Phys. Rev. Lett.}\ }\textbf {\bibinfo {volume} {127}},\ \bibinfo {pages} {068001} (\bibinfo {year} {2021})}\BibitemShut {NoStop}%
\bibitem [{\citenamefont {Caballero}\ and\ \citenamefont {Marchetti}(2022)}]{caballero2022activity}%
  \BibitemOpen
  \bibfield  {author} {\bibinfo {author} {\bibfnamefont {F.}~\bibnamefont {Caballero}}\ and\ \bibinfo {author} {\bibfnamefont {M.~C.}\ \bibnamefont {Marchetti}},\ }\href@noop {} {\bibfield  {journal} {\bibinfo  {journal} {Phys. Rev. Lett.}\ }\textbf {\bibinfo {volume} {129}},\ \bibinfo {pages} {268002} (\bibinfo {year} {2022})}\BibitemShut {NoStop}%
\bibitem [{\citenamefont {Gulati}\ \emph {et~al.}(2024)\citenamefont {Gulati}, \citenamefont {Caballero}, \citenamefont {Kolvin}, \citenamefont {You},\ and\ \citenamefont {Marchetti}}]{gulati2024nonreciprocal}%
  \BibitemOpen
  \bibfield  {author} {\bibinfo {author} {\bibfnamefont {P.}~\bibnamefont {Gulati}}, \bibinfo {author} {\bibfnamefont {F.}~\bibnamefont {Caballero}}, \bibinfo {author} {\bibfnamefont {I.}~\bibnamefont {Kolvin}}, \bibinfo {author} {\bibfnamefont {Z.}~\bibnamefont {You}}, \ and\ \bibinfo {author} {\bibfnamefont {M.~C.}\ \bibnamefont {Marchetti}},\ }\href@noop {} {\bibfield  {journal} {\bibinfo  {journal} {arXiv preprint arXiv:2407.04196}\ } (\bibinfo {year} {2024})}\BibitemShut {NoStop}%
\bibitem [{\citenamefont {Besse}\ \emph {et~al.}(2023)\citenamefont {Besse}, \citenamefont {Fausti}, \citenamefont {Cates}, \citenamefont {Delamotte},\ and\ \citenamefont {Nardini}}]{besse2023interface}%
  \BibitemOpen
  \bibfield  {author} {\bibinfo {author} {\bibfnamefont {M.}~\bibnamefont {Besse}}, \bibinfo {author} {\bibfnamefont {G.}~\bibnamefont {Fausti}}, \bibinfo {author} {\bibfnamefont {M.~E.}\ \bibnamefont {Cates}}, \bibinfo {author} {\bibfnamefont {B.}~\bibnamefont {Delamotte}}, \ and\ \bibinfo {author} {\bibfnamefont {C.}~\bibnamefont {Nardini}},\ }\href@noop {} {\bibfield  {journal} {\bibinfo  {journal} {Phys. Rev. Lett.}\ }\textbf {\bibinfo {volume} {130}},\ \bibinfo {pages} {187102} (\bibinfo {year} {2023})}\BibitemShut {NoStop}%
\bibitem [{\citenamefont {Marchetti}\ \emph {et~al.}(2013)\citenamefont {Marchetti}, \citenamefont {Joanny}, \citenamefont {Ramaswamy}, \citenamefont {Liverpool}, \citenamefont {Prost}, \citenamefont {Rao},\ and\ \citenamefont {Simha}}]{marchetti2013hydrodynamics}%
  \BibitemOpen
  \bibfield  {author} {\bibinfo {author} {\bibfnamefont {M.~C.}\ \bibnamefont {Marchetti}}, \bibinfo {author} {\bibfnamefont {J.-F.}\ \bibnamefont {Joanny}}, \bibinfo {author} {\bibfnamefont {S.}~\bibnamefont {Ramaswamy}}, \bibinfo {author} {\bibfnamefont {T.~B.}\ \bibnamefont {Liverpool}}, \bibinfo {author} {\bibfnamefont {J.}~\bibnamefont {Prost}}, \bibinfo {author} {\bibfnamefont {M.}~\bibnamefont {Rao}}, \ and\ \bibinfo {author} {\bibfnamefont {R.~A.}\ \bibnamefont {Simha}},\ }\href@noop {} {\bibfield  {journal} {\bibinfo  {journal} {Rev. Mod. Phys.}\ }\textbf {\bibinfo {volume} {85}},\ \bibinfo {pages} {1143} (\bibinfo {year} {2013})}\BibitemShut {NoStop}%
\bibitem [{\citenamefont {J{\"u}licher}\ \emph {et~al.}(2018)\citenamefont {J{\"u}licher}, \citenamefont {Grill},\ and\ \citenamefont {Salbreux}}]{julicher2018hydrodynamic}%
  \BibitemOpen
  \bibfield  {author} {\bibinfo {author} {\bibfnamefont {F.}~\bibnamefont {J{\"u}licher}}, \bibinfo {author} {\bibfnamefont {S.~W.}\ \bibnamefont {Grill}}, \ and\ \bibinfo {author} {\bibfnamefont {G.}~\bibnamefont {Salbreux}},\ }\href@noop {} {\bibfield  {journal} {\bibinfo  {journal} {Rep. Prog. Phys.}\ }\textbf {\bibinfo {volume} {81}},\ \bibinfo {pages} {076601} (\bibinfo {year} {2018})}\BibitemShut {NoStop}%
\bibitem [{\citenamefont {Palacci}\ \emph {et~al.}(2013)\citenamefont {Palacci}, \citenamefont {Sacanna}, \citenamefont {Steinberg}, \citenamefont {Pine},\ and\ \citenamefont {Chaikin}}]{palacci2013living}%
  \BibitemOpen
  \bibfield  {author} {\bibinfo {author} {\bibfnamefont {J.}~\bibnamefont {Palacci}}, \bibinfo {author} {\bibfnamefont {S.}~\bibnamefont {Sacanna}}, \bibinfo {author} {\bibfnamefont {A.~P.}\ \bibnamefont {Steinberg}}, \bibinfo {author} {\bibfnamefont {D.~J.}\ \bibnamefont {Pine}}, \ and\ \bibinfo {author} {\bibfnamefont {P.~M.}\ \bibnamefont {Chaikin}},\ }\href@noop {} {\bibfield  {journal} {\bibinfo  {journal} {Science}\ }\textbf {\bibinfo {volume} {339}},\ \bibinfo {pages} {936} (\bibinfo {year} {2013})}\BibitemShut {NoStop}%
\bibitem [{\citenamefont {Sanchez}\ \emph {et~al.}(2012)\citenamefont {Sanchez}, \citenamefont {Chen}, \citenamefont {DeCamp}, \citenamefont {Heymann},\ and\ \citenamefont {Dogic}}]{sanchez2012spontaneous}%
  \BibitemOpen
  \bibfield  {author} {\bibinfo {author} {\bibfnamefont {T.}~\bibnamefont {Sanchez}}, \bibinfo {author} {\bibfnamefont {D.~T.}\ \bibnamefont {Chen}}, \bibinfo {author} {\bibfnamefont {S.~J.}\ \bibnamefont {DeCamp}}, \bibinfo {author} {\bibfnamefont {M.}~\bibnamefont {Heymann}}, \ and\ \bibinfo {author} {\bibfnamefont {Z.}~\bibnamefont {Dogic}},\ }\href@noop {} {\bibfield  {journal} {\bibinfo  {journal} {Nature}\ }\textbf {\bibinfo {volume} {491}},\ \bibinfo {pages} {431} (\bibinfo {year} {2012})}\BibitemShut {NoStop}%
\bibitem [{\citenamefont {Bricard}\ \emph {et~al.}(2013{\natexlab{a}})\citenamefont {Bricard}, \citenamefont {Caussin}, \citenamefont {Desreumaux}, \citenamefont {Dauchot},\ and\ \citenamefont {Bartolo}}]{bricard2013emergence}%
  \BibitemOpen
  \bibfield  {author} {\bibinfo {author} {\bibfnamefont {A.}~\bibnamefont {Bricard}}, \bibinfo {author} {\bibfnamefont {J.-B.}\ \bibnamefont {Caussin}}, \bibinfo {author} {\bibfnamefont {N.}~\bibnamefont {Desreumaux}}, \bibinfo {author} {\bibfnamefont {O.}~\bibnamefont {Dauchot}}, \ and\ \bibinfo {author} {\bibfnamefont {D.}~\bibnamefont {Bartolo}},\ }\href@noop {} {\bibfield  {journal} {\bibinfo  {journal} {Nature}\ }\textbf {\bibinfo {volume} {503}},\ \bibinfo {pages} {95} (\bibinfo {year} {2013}{\natexlab{a}})}\BibitemShut {NoStop}%
\bibitem [{\citenamefont {Patteson}\ \emph {et~al.}(2018)\citenamefont {Patteson}, \citenamefont {Gopinath},\ and\ \citenamefont {Arratia}}]{patteson2018propagation}%
  \BibitemOpen
  \bibfield  {author} {\bibinfo {author} {\bibfnamefont {A.~E.}\ \bibnamefont {Patteson}}, \bibinfo {author} {\bibfnamefont {A.}~\bibnamefont {Gopinath}}, \ and\ \bibinfo {author} {\bibfnamefont {P.~E.}\ \bibnamefont {Arratia}},\ }\href@noop {} {\bibfield  {journal} {\bibinfo  {journal} {Nat. Comm.}\ }\textbf {\bibinfo {volume} {9}},\ \bibinfo {pages} {5373} (\bibinfo {year} {2018})}\BibitemShut {NoStop}%
\bibitem [{\citenamefont {Adkins}\ \emph {et~al.}(2022)\citenamefont {Adkins}, \citenamefont {Kolvin}, \citenamefont {You}, \citenamefont {Witthaus}, \citenamefont {Marchetti},\ and\ \citenamefont {Dogic}}]{adkins2022dynamics}%
  \BibitemOpen
  \bibfield  {author} {\bibinfo {author} {\bibfnamefont {R.}~\bibnamefont {Adkins}}, \bibinfo {author} {\bibfnamefont {I.}~\bibnamefont {Kolvin}}, \bibinfo {author} {\bibfnamefont {Z.}~\bibnamefont {You}}, \bibinfo {author} {\bibfnamefont {S.}~\bibnamefont {Witthaus}}, \bibinfo {author} {\bibfnamefont {M.~C.}\ \bibnamefont {Marchetti}}, \ and\ \bibinfo {author} {\bibfnamefont {Z.}~\bibnamefont {Dogic}},\ }\href@noop {} {\bibfield  {journal} {\bibinfo  {journal} {Science}\ }\textbf {\bibinfo {volume} {377}},\ \bibinfo {pages} {768} (\bibinfo {year} {2022})}\BibitemShut {NoStop}%
\bibitem [{\citenamefont {Tayar}\ \emph {et~al.}(2023)\citenamefont {Tayar}, \citenamefont {Caballero}, \citenamefont {Anderberg}, \citenamefont {Saleh}, \citenamefont {Cristina~Marchetti},\ and\ \citenamefont {Dogic}}]{tayar2023controlling}%
  \BibitemOpen
  \bibfield  {author} {\bibinfo {author} {\bibfnamefont {A.~M.}\ \bibnamefont {Tayar}}, \bibinfo {author} {\bibfnamefont {F.}~\bibnamefont {Caballero}}, \bibinfo {author} {\bibfnamefont {T.}~\bibnamefont {Anderberg}}, \bibinfo {author} {\bibfnamefont {O.~A.}\ \bibnamefont {Saleh}}, \bibinfo {author} {\bibfnamefont {M.}~\bibnamefont {Cristina~Marchetti}}, \ and\ \bibinfo {author} {\bibfnamefont {Z.}~\bibnamefont {Dogic}},\ }\href@noop {} {\bibfield  {journal} {\bibinfo  {journal} {Nat. Mater.}\ }\textbf {\bibinfo {volume} {22}},\ \bibinfo {pages} {1401} (\bibinfo {year} {2023})}\BibitemShut {NoStop}%
\bibitem [{\citenamefont {Galajda}\ \emph {et~al.}(2007)\citenamefont {Galajda}, \citenamefont {Keymer}, \citenamefont {Chaikin},\ and\ \citenamefont {Austin}}]{galajda2007wall}%
  \BibitemOpen
  \bibfield  {author} {\bibinfo {author} {\bibfnamefont {P.}~\bibnamefont {Galajda}}, \bibinfo {author} {\bibfnamefont {J.}~\bibnamefont {Keymer}}, \bibinfo {author} {\bibfnamefont {P.}~\bibnamefont {Chaikin}}, \ and\ \bibinfo {author} {\bibfnamefont {R.}~\bibnamefont {Austin}},\ }\href@noop {} {\bibfield  {journal} {\bibinfo  {journal} {J. Bact.}\ }\textbf {\bibinfo {volume} {189}},\ \bibinfo {pages} {8704} (\bibinfo {year} {2007})}\BibitemShut {NoStop}%
\bibitem [{\citenamefont {Diamant}(2009)}]{diamant2009hydrodynamic}%
  \BibitemOpen
  \bibfield  {author} {\bibinfo {author} {\bibfnamefont {H.}~\bibnamefont {Diamant}},\ }\href@noop {} {\bibfield  {journal} {\bibinfo  {journal} {J. Phys. Soc. Jpn}\ }\textbf {\bibinfo {volume} {78}},\ \bibinfo {pages} {041002} (\bibinfo {year} {2009})}\BibitemShut {NoStop}%
\bibitem [{\citenamefont {Maitra}\ \emph {et~al.}(2020)\citenamefont {Maitra}, \citenamefont {Srivastava}, \citenamefont {Marchetti}, \citenamefont {Ramaswamy},\ and\ \citenamefont {Lenz}}]{maitra2020swimmer}%
  \BibitemOpen
  \bibfield  {author} {\bibinfo {author} {\bibfnamefont {A.}~\bibnamefont {Maitra}}, \bibinfo {author} {\bibfnamefont {P.}~\bibnamefont {Srivastava}}, \bibinfo {author} {\bibfnamefont {M.~C.}\ \bibnamefont {Marchetti}}, \bibinfo {author} {\bibfnamefont {S.}~\bibnamefont {Ramaswamy}}, \ and\ \bibinfo {author} {\bibfnamefont {M.}~\bibnamefont {Lenz}},\ }\href@noop {} {\bibfield  {journal} {\bibinfo  {journal} {Phys. Rev. Lett.}\ }\textbf {\bibinfo {volume} {124}},\ \bibinfo {pages} {028002} (\bibinfo {year} {2020})}\BibitemShut {NoStop}%
\bibitem [{\citenamefont {Chen}\ \emph {et~al.}(2016)\citenamefont {Chen}, \citenamefont {Lee},\ and\ \citenamefont {Toner}}]{CLT_Ncom}%
  \BibitemOpen
  \bibfield  {author} {\bibinfo {author} {\bibfnamefont {L.}~\bibnamefont {Chen}}, \bibinfo {author} {\bibfnamefont {C.~F.}\ \bibnamefont {Lee}}, \ and\ \bibinfo {author} {\bibfnamefont {J.}~\bibnamefont {Toner}},\ }\href@noop {} {\bibfield  {journal} {\bibinfo  {journal} {Nat. Comm.}\ }\textbf {\bibinfo {volume} {7}},\ \bibinfo {pages} {12215} (\bibinfo {year} {2016})}\BibitemShut {NoStop}%
\bibitem [{\citenamefont {Bricard}\ \emph {et~al.}(2013{\natexlab{b}})\citenamefont {Bricard}, \citenamefont {Caussin}, \citenamefont {Desreumaux}, \citenamefont {Dauchot},\ and\ \citenamefont {Bartolo}}]{Bricard}%
  \BibitemOpen
  \bibfield  {author} {\bibinfo {author} {\bibfnamefont {A.}~\bibnamefont {Bricard}}, \bibinfo {author} {\bibfnamefont {J.-B.}\ \bibnamefont {Caussin}}, \bibinfo {author} {\bibfnamefont {N.}~\bibnamefont {Desreumaux}}, \bibinfo {author} {\bibfnamefont {O.}~\bibnamefont {Dauchot}}, \ and\ \bibinfo {author} {\bibfnamefont {D.}~\bibnamefont {Bartolo}},\ }\href@noop {} {\bibfield  {journal} {\bibinfo  {journal} {Nature}\ }\textbf {\bibinfo {volume} {503}},\ \bibinfo {pages} {95} (\bibinfo {year} {2013}{\natexlab{b}})}\BibitemShut {NoStop}%
\bibitem [{\citenamefont {Brotto}\ \emph {et~al.}(2013)\citenamefont {Brotto}, \citenamefont {Caussin}, \citenamefont {Lauga},\ and\ \citenamefont {Bartolo}}]{Brotto}%
  \BibitemOpen
  \bibfield  {author} {\bibinfo {author} {\bibfnamefont {T.}~\bibnamefont {Brotto}}, \bibinfo {author} {\bibfnamefont {J.-B.}\ \bibnamefont {Caussin}}, \bibinfo {author} {\bibfnamefont {E.}~\bibnamefont {Lauga}}, \ and\ \bibinfo {author} {\bibfnamefont {D.}~\bibnamefont {Bartolo}},\ }\href@noop {} {\bibfield  {journal} {\bibinfo  {journal} {Phys. Rev. Lett.}\ }\textbf {\bibinfo {volume} {110}},\ \bibinfo {pages} {038101} (\bibinfo {year} {2013})}\BibitemShut {NoStop}%
\bibitem [{\citenamefont {Sarkar}\ \emph {et~al.}(2021)\citenamefont {Sarkar}, \citenamefont {Basu},\ and\ \citenamefont {Toner}}]{Sarkar_Toner}%
  \BibitemOpen
  \bibfield  {author} {\bibinfo {author} {\bibfnamefont {N.}~\bibnamefont {Sarkar}}, \bibinfo {author} {\bibfnamefont {A.}~\bibnamefont {Basu}}, \ and\ \bibinfo {author} {\bibfnamefont {J.}~\bibnamefont {Toner}},\ }\href@noop {} {\bibfield  {journal} {\bibinfo  {journal} {Phys. Rev. Lett.}\ }\textbf {\bibinfo {volume} {127}},\ \bibinfo {pages} {268004} (\bibinfo {year} {2021})}\BibitemShut {NoStop}%
\bibitem [{\citenamefont {Cates}\ and\ \citenamefont {Tjhung}(2018)}]{cates2018theories}%
  \BibitemOpen
  \bibfield  {author} {\bibinfo {author} {\bibfnamefont {M.~E.}\ \bibnamefont {Cates}}\ and\ \bibinfo {author} {\bibfnamefont {E.}~\bibnamefont {Tjhung}},\ }\href@noop {} {\bibfield  {journal} {\bibinfo  {journal} {J. Fluid Mech.}\ }\textbf {\bibinfo {volume} {836}},\ \bibinfo {pages} {P1} (\bibinfo {year} {2018})}\BibitemShut {NoStop}%
\bibitem [{\citenamefont {Hohenberg}\ and\ \citenamefont {Halperin}(1977)}]{hohenberg1977theory}%
  \BibitemOpen
  \bibfield  {author} {\bibinfo {author} {\bibfnamefont {P.~C.}\ \bibnamefont {Hohenberg}}\ and\ \bibinfo {author} {\bibfnamefont {B.~I.}\ \bibnamefont {Halperin}},\ }\href@noop {} {\bibfield  {journal} {\bibinfo  {journal} {Rev. Mod. Phys.}\ }\textbf {\bibinfo {volume} {49}},\ \bibinfo {pages} {435} (\bibinfo {year} {1977})}\BibitemShut {NoStop}%
\bibitem [{\citenamefont {Oron}\ \emph {et~al.}(1997)\citenamefont {Oron}, \citenamefont {Davis},\ and\ \citenamefont {Bankoff}}]{Oron}%
  \BibitemOpen
  \bibfield  {author} {\bibinfo {author} {\bibfnamefont {A.}~\bibnamefont {Oron}}, \bibinfo {author} {\bibfnamefont {S.~H.}\ \bibnamefont {Davis}}, \ and\ \bibinfo {author} {\bibfnamefont {S.~G.}\ \bibnamefont {Bankoff}},\ }\href@noop {} {\bibfield  {journal} {\bibinfo  {journal} {Rev. Mod. Phys.}\ }\textbf {\bibinfo {volume} {69}},\ \bibinfo {pages} {931} (\bibinfo {year} {1997})}\BibitemShut {NoStop}%
\bibitem [{\citenamefont {Maitra}\ \emph {et~al.}(2018)\citenamefont {Maitra}, \citenamefont {Srivastava}, \citenamefont {Marchetti}, \citenamefont {Lintuvuori}, \citenamefont {Ramaswamy},\ and\ \citenamefont {Lenz}}]{AM_nem}%
  \BibitemOpen
  \bibfield  {author} {\bibinfo {author} {\bibfnamefont {A.}~\bibnamefont {Maitra}}, \bibinfo {author} {\bibfnamefont {P.}~\bibnamefont {Srivastava}}, \bibinfo {author} {\bibfnamefont {M.~C.}\ \bibnamefont {Marchetti}}, \bibinfo {author} {\bibfnamefont {J.~S.}\ \bibnamefont {Lintuvuori}}, \bibinfo {author} {\bibfnamefont {S.}~\bibnamefont {Ramaswamy}}, \ and\ \bibinfo {author} {\bibfnamefont {M.}~\bibnamefont {Lenz}},\ }\href@noop {} {\bibfield  {journal} {\bibinfo  {journal} {Proc. Nat. A. Sci.}\ }\textbf {\bibinfo {volume} {115}},\ \bibinfo {pages} {6934} (\bibinfo {year} {2018})}\BibitemShut {NoStop}%
\bibitem [{\citenamefont {Tiribocchi}\ \emph {et~al.}(2015)\citenamefont {Tiribocchi}, \citenamefont {Wittkowski}, \citenamefont {Marenduzzo},\ and\ \citenamefont {Cates}}]{tiribocchi2015active}%
  \BibitemOpen
  \bibfield  {author} {\bibinfo {author} {\bibfnamefont {A.}~\bibnamefont {Tiribocchi}}, \bibinfo {author} {\bibfnamefont {R.}~\bibnamefont {Wittkowski}}, \bibinfo {author} {\bibfnamefont {D.}~\bibnamefont {Marenduzzo}}, \ and\ \bibinfo {author} {\bibfnamefont {M.~E.}\ \bibnamefont {Cates}},\ }\href@noop {} {\bibfield  {journal} {\bibinfo  {journal} {Phys. Rev. Lett.}\ }\textbf {\bibinfo {volume} {115}},\ \bibinfo {pages} {188302} (\bibinfo {year} {2015})}\BibitemShut {NoStop}%
\bibitem [{\citenamefont {Prost}\ and\ \citenamefont {Bruinsma}(1996)}]{prost1996shape}%
  \BibitemOpen
  \bibfield  {author} {\bibinfo {author} {\bibfnamefont {J.}~\bibnamefont {Prost}}\ and\ \bibinfo {author} {\bibfnamefont {R.}~\bibnamefont {Bruinsma}},\ }\href@noop {} {\bibfield  {journal} {\bibinfo  {journal} {Europhysics Letters}\ }\textbf {\bibinfo {volume} {33}},\ \bibinfo {pages} {321} (\bibinfo {year} {1996})}\BibitemShut {NoStop}%
\bibitem [{\citenamefont {Cagnetta}\ \emph {et~al.}(2022)\citenamefont {Cagnetta}, \citenamefont {{\v{S}}kult{\'e}ty}, \citenamefont {Evans},\ and\ \citenamefont {Marenduzzo}}]{cagnetta2022universal}%
  \BibitemOpen
  \bibfield  {author} {\bibinfo {author} {\bibfnamefont {F.}~\bibnamefont {Cagnetta}}, \bibinfo {author} {\bibfnamefont {V.}~\bibnamefont {{\v{S}}kult{\'e}ty}}, \bibinfo {author} {\bibfnamefont {M.~R.}\ \bibnamefont {Evans}}, \ and\ \bibinfo {author} {\bibfnamefont {D.}~\bibnamefont {Marenduzzo}},\ }\href@noop {} {\bibfield  {journal} {\bibinfo  {journal} {Physical Review E}\ }\textbf {\bibinfo {volume} {105}},\ \bibinfo {pages} {L012604} (\bibinfo {year} {2022})}\BibitemShut {NoStop}%
\bibitem [{\citenamefont {Ramaswamy}\ \emph {et~al.}(2000)\citenamefont {Ramaswamy}, \citenamefont {Toner},\ and\ \citenamefont {Prost}}]{ramaswamy2000nonequilibrium}%
  \BibitemOpen
  \bibfield  {author} {\bibinfo {author} {\bibfnamefont {S.}~\bibnamefont {Ramaswamy}}, \bibinfo {author} {\bibfnamefont {J.}~\bibnamefont {Toner}}, \ and\ \bibinfo {author} {\bibfnamefont {J.}~\bibnamefont {Prost}},\ }\href@noop {} {\bibfield  {journal} {\bibinfo  {journal} {Physical review letters}\ }\textbf {\bibinfo {volume} {84}},\ \bibinfo {pages} {3494} (\bibinfo {year} {2000})}\BibitemShut {NoStop}%
\bibitem [{foo()}]{footnote}%
  \BibitemOpen
  \href@noop {} {}\bibinfo {note} {Notice that it has a non-zero value in the steady state $2\alpha\delta(y)\hat{y}$. It can be shown that this leads only to a pressure jump at the interface and not to a steady motion.}\BibitemShut {Stop}%
\bibitem [{SM()}]{SM}%
  \BibitemOpen
  \href@noop {} {}\bibinfo {note} {See Supplemental Material at [URL will be inserted by publisher] which presents additional numerical results and details.}\BibitemShut {Stop}%
\bibitem [{\citenamefont {Caballero}(2024)}]{caballero2024cupss}%
  \BibitemOpen
  \bibfield  {author} {\bibinfo {author} {\bibfnamefont {F.}~\bibnamefont {Caballero}},\ }\href@noop {} {\bibfield  {journal} {\bibinfo  {journal} {arXiv preprint arXiv:2405.02410}\ } (\bibinfo {year} {2024})}\BibitemShut {NoStop}%
\bibitem [{\citenamefont {De~Gennes}\ and\ \citenamefont {Taupin}(1982)}]{Taupin}%
  \BibitemOpen
  \bibfield  {author} {\bibinfo {author} {\bibfnamefont {P.}~\bibnamefont {De~Gennes}}\ and\ \bibinfo {author} {\bibfnamefont {C.}~\bibnamefont {Taupin}},\ }\href@noop {} {\bibfield  {journal} {\bibinfo  {journal} {J. Phys. Chem.}\ }\textbf {\bibinfo {volume} {86}},\ \bibinfo {pages} {2294} (\bibinfo {year} {1982})}\BibitemShut {NoStop}%
\bibitem [{\citenamefont {Sussman}\ \emph {et~al.}(2018)\citenamefont {Sussman}, \citenamefont {Schwarz}, \citenamefont {Marchetti},\ and\ \citenamefont {Manning}}]{sussman1}%
  \BibitemOpen
  \bibfield  {author} {\bibinfo {author} {\bibfnamefont {D.~M.}\ \bibnamefont {Sussman}}, \bibinfo {author} {\bibfnamefont {J.}~\bibnamefont {Schwarz}}, \bibinfo {author} {\bibfnamefont {M.~C.}\ \bibnamefont {Marchetti}}, \ and\ \bibinfo {author} {\bibfnamefont {M.~L.}\ \bibnamefont {Manning}},\ }\href@noop {} {\bibfield  {journal} {\bibinfo  {journal} {Phys. Rev. Lett.}\ }\textbf {\bibinfo {volume} {120}},\ \bibinfo {pages} {058001} (\bibinfo {year} {2018})}\BibitemShut {NoStop}%
\bibitem [{\citenamefont {Yue}\ \emph {et~al.}(2024)\citenamefont {Yue}, \citenamefont {Packard},\ and\ \citenamefont {Sussman}}]{sussman2}%
  \BibitemOpen
  \bibfield  {author} {\bibinfo {author} {\bibfnamefont {H.}~\bibnamefont {Yue}}, \bibinfo {author} {\bibfnamefont {C.~R.}\ \bibnamefont {Packard}}, \ and\ \bibinfo {author} {\bibfnamefont {D.~M.}\ \bibnamefont {Sussman}},\ }\href@noop {} {\bibfield  {journal} {\bibinfo  {journal} {arXiv preprint arXiv:2407.02760}\ } (\bibinfo {year} {2024})}\BibitemShut {NoStop}%
\bibitem [{\citenamefont {Zhao}\ \emph {et~al.}(2024)\citenamefont {Zhao}, \citenamefont {Gulati}, \citenamefont {Caballero}, \citenamefont {Kolvin}, \citenamefont {Adkins}, \citenamefont {Marchetti},\ and\ \citenamefont {Dogic}}]{zhao2024asymmetric}%
  \BibitemOpen
  \bibfield  {author} {\bibinfo {author} {\bibfnamefont {L.}~\bibnamefont {Zhao}}, \bibinfo {author} {\bibfnamefont {P.}~\bibnamefont {Gulati}}, \bibinfo {author} {\bibfnamefont {F.}~\bibnamefont {Caballero}}, \bibinfo {author} {\bibfnamefont {I.}~\bibnamefont {Kolvin}}, \bibinfo {author} {\bibfnamefont {R.}~\bibnamefont {Adkins}}, \bibinfo {author} {\bibfnamefont {M.~C.}\ \bibnamefont {Marchetti}}, \ and\ \bibinfo {author} {\bibfnamefont {Z.}~\bibnamefont {Dogic}},\ }\href@noop {} {\bibfield  {journal} {\bibinfo  {journal} {arXiv preprint arXiv:2407.04679}\ } (\bibinfo {year} {2024})}\BibitemShut {NoStop}%
\bibitem [{\citenamefont {Law}\ \emph {et~al.}(2023)\citenamefont {Law}, \citenamefont {Jones}, \citenamefont {Stevenson}, \citenamefont {Williamson}, \citenamefont {Turner}, \citenamefont {Kusumaatmaja},\ and\ \citenamefont {Grellscheid}}]{law2023bending}%
  \BibitemOpen
  \bibfield  {author} {\bibinfo {author} {\bibfnamefont {J.~O.}\ \bibnamefont {Law}}, \bibinfo {author} {\bibfnamefont {C.~M.}\ \bibnamefont {Jones}}, \bibinfo {author} {\bibfnamefont {T.}~\bibnamefont {Stevenson}}, \bibinfo {author} {\bibfnamefont {T.~A.}\ \bibnamefont {Williamson}}, \bibinfo {author} {\bibfnamefont {M.~S.}\ \bibnamefont {Turner}}, \bibinfo {author} {\bibfnamefont {H.}~\bibnamefont {Kusumaatmaja}}, \ and\ \bibinfo {author} {\bibfnamefont {S.~N.}\ \bibnamefont {Grellscheid}},\ }\href@noop {} {\bibfield  {journal} {\bibinfo  {journal} {Science Advances}\ }\textbf {\bibinfo {volume} {9}},\ \bibinfo {pages} {eadg0432} (\bibinfo {year} {2023})}\BibitemShut {NoStop}%
\bibitem [{\citenamefont {Caragine}\ \emph {et~al.}(2018)\citenamefont {Caragine}, \citenamefont {Haley},\ and\ \citenamefont {Zidovska}}]{caragine2018surface}%
  \BibitemOpen
  \bibfield  {author} {\bibinfo {author} {\bibfnamefont {C.~M.}\ \bibnamefont {Caragine}}, \bibinfo {author} {\bibfnamefont {S.~C.}\ \bibnamefont {Haley}}, \ and\ \bibinfo {author} {\bibfnamefont {A.}~\bibnamefont {Zidovska}},\ }\href@noop {} {\bibfield  {journal} {\bibinfo  {journal} {Physical review letters}\ }\textbf {\bibinfo {volume} {121}},\ \bibinfo {pages} {148101} (\bibinfo {year} {2018})}\BibitemShut {NoStop}%
\bibitem [{\citenamefont {Seyboldt}\ and\ \citenamefont {J{\"u}licher}(2018)}]{seyboldt2018role}%
  \BibitemOpen
  \bibfield  {author} {\bibinfo {author} {\bibfnamefont {R.}~\bibnamefont {Seyboldt}}\ and\ \bibinfo {author} {\bibfnamefont {F.}~\bibnamefont {J{\"u}licher}},\ }\href@noop {} {\bibfield  {journal} {\bibinfo  {journal} {New journal of physics}\ }\textbf {\bibinfo {volume} {20}},\ \bibinfo {pages} {105010} (\bibinfo {year} {2018})}\BibitemShut {NoStop}%
\bibitem [{\citenamefont {Galvanetto}\ \emph {et~al.}(2023)\citenamefont {Galvanetto}, \citenamefont {Ivanovi{\'c}}, \citenamefont {Chowdhury}, \citenamefont {Sottini}, \citenamefont {N{\"u}esch}, \citenamefont {Nettels}, \citenamefont {Best},\ and\ \citenamefont {Schuler}}]{galvanetto2023extreme}%
  \BibitemOpen
  \bibfield  {author} {\bibinfo {author} {\bibfnamefont {N.}~\bibnamefont {Galvanetto}}, \bibinfo {author} {\bibfnamefont {M.~T.}\ \bibnamefont {Ivanovi{\'c}}}, \bibinfo {author} {\bibfnamefont {A.}~\bibnamefont {Chowdhury}}, \bibinfo {author} {\bibfnamefont {A.}~\bibnamefont {Sottini}}, \bibinfo {author} {\bibfnamefont {M.~F.}\ \bibnamefont {N{\"u}esch}}, \bibinfo {author} {\bibfnamefont {D.}~\bibnamefont {Nettels}}, \bibinfo {author} {\bibfnamefont {R.~B.}\ \bibnamefont {Best}}, \ and\ \bibinfo {author} {\bibfnamefont {B.}~\bibnamefont {Schuler}},\ }\href@noop {} {\bibfield  {journal} {\bibinfo  {journal} {Nature}\ }\textbf {\bibinfo {volume} {619}},\ \bibinfo {pages} {876} (\bibinfo {year} {2023})}\BibitemShut {NoStop}%
\end{thebibliography}%


\begin{thebibliography}{3}%
\makeatletter
\providecommand \@ifxundefined [1]{%
 \@ifx{#1\undefined}
}%
\providecommand \@ifnum [1]{%
 \ifnum #1\expandafter \@firstoftwo
 \else \expandafter \@secondoftwo
 \fi
}%
\providecommand \@ifx [1]{%
 \ifx #1\expandafter \@firstoftwo
 \else \expandafter \@secondoftwo
 \fi
}%
\providecommand \natexlab [1]{#1}%
\providecommand \enquote  [1]{``#1''}%
\providecommand \bibnamefont  [1]{#1}%
\providecommand \bibfnamefont [1]{#1}%
\providecommand \citenamefont [1]{#1}%
\providecommand \href@noop [0]{\@secondoftwo}%
\providecommand \href [0]{\begingroup \@sanitize@url \@href}%
\providecommand \@href[1]{\@@startlink{#1}\@@href}%
\providecommand \@@href[1]{\endgroup#1\@@endlink}%
\providecommand \@sanitize@url [0]{\catcode `\\12\catcode `\$12\catcode `\&12\catcode `\#12\catcode `\^12\catcode `\_12\catcode `\%12\relax}%
\providecommand \@@startlink[1]{}%
\providecommand \@@endlink[0]{}%
\providecommand \url  [0]{\begingroup\@sanitize@url \@url }%
\providecommand \@url [1]{\endgroup\@href {#1}{\urlprefix }}%
\providecommand \urlprefix  [0]{URL }%
\providecommand \Eprint [0]{\href }%
\providecommand \doibase [0]{https://doi.org/}%
\providecommand \selectlanguage [0]{\@gobble}%
\providecommand \bibinfo  [0]{\@secondoftwo}%
\providecommand \bibfield  [0]{\@secondoftwo}%
\providecommand \translation [1]{[#1]}%
\providecommand \BibitemOpen [0]{}%
\providecommand \bibitemStop [0]{}%
\providecommand \bibitemNoStop [0]{.\EOS\space}%
\providecommand \EOS [0]{\spacefactor3000\relax}%
\providecommand \BibitemShut  [1]{\csname bibitem#1\endcsname}%
\let\auto@bib@innerbib\@empty
\bibitem [{\citenamefont {Caballero}(2024)}]{caballero2024cupss}%
  \BibitemOpen
  \bibfield  {author} {\bibinfo {author} {\bibfnamefont {F.}~\bibnamefont {Caballero}},\ }\bibfield  {title} {\bibinfo {title} {cu{PSS}: a package for pseudo-spectral integration of stochastic {PDE}s},\ }\href@noop {} {\bibfield  {journal} {\bibinfo  {journal} {arXiv preprint arXiv:2405.02410}\ } (\bibinfo {year} {2024})}\BibitemShut {NoStop}%
\bibitem [{\citenamefont {Bray}\ \emph {et~al.}(2001)\citenamefont {Bray}, \citenamefont {Cavagna},\ and\ \citenamefont {Travasso}}]{bray2001interface}%
  \BibitemOpen
  \bibfield  {author} {\bibinfo {author} {\bibfnamefont {A.~J.}\ \bibnamefont {Bray}}, \bibinfo {author} {\bibfnamefont {A.}~\bibnamefont {Cavagna}},\ and\ \bibinfo {author} {\bibfnamefont {R.~D.}\ \bibnamefont {Travasso}},\ }\bibfield  {title} {\bibinfo {title} {Interface fluctuations, burgers equations, and coarsening under shear},\ }\href@noop {} {\bibfield  {journal} {\bibinfo  {journal} {Phys. Rev. E}\ }\textbf {\bibinfo {volume} {65}},\ \bibinfo {pages} {016104} (\bibinfo {year} {2001})}\BibitemShut {NoStop}%
\bibitem [{\citenamefont {Kardar}\ \emph {et~al.}(1986)\citenamefont {Kardar}, \citenamefont {Parisi},\ and\ \citenamefont {Zhang}}]{kardar1986dynamic}%
  \BibitemOpen
  \bibfield  {author} {\bibinfo {author} {\bibfnamefont {M.}~\bibnamefont {Kardar}}, \bibinfo {author} {\bibfnamefont {G.}~\bibnamefont {Parisi}},\ and\ \bibinfo {author} {\bibfnamefont {Y.-C.}\ \bibnamefont {Zhang}},\ }\bibfield  {title} {\bibinfo {title} {Dynamic scaling of growing interfaces},\ }\href@noop {} {\bibfield  {journal} {\bibinfo  {journal} {Phys. Rev. Lett.}\ }\textbf {\bibinfo {volume} {56}},\ \bibinfo {pages} {889} (\bibinfo {year} {1986})}\BibitemShut {NoStop}%
\end{thebibliography}%










\end{document}


\preprint{APS/123-QED}

\title{Supplementary Information: Interface dynamics of wet active systems}

\author{Fernando Caballero}
 \affiliation{Department of Physics, Brandeis University, Waltham, MA 02453, USA}
 \email{fcaballero@brandeis.edu}
\author{Ananyo Maitra}%
\affiliation{Laboratoire de Physique Th\'eorique et Mod\'elisation, CNRS UMR 8089, CY Cergy Paris Universit\'e, F-95032 Cergy-Pontoise Cedex, France}
\affiliation{Laboratoire Jean Perrin, UMR 8237 CNRS, Sorbonne Universit\'e, 75005 Paris, France}
\email{nyomaitra07@gmail.com}
\author{Cesare Nardini}
\affiliation{Service de Physique de l'Etat Condens\'e, CEA, CNRS Universit\'e Paris-Saclay, CEA-Saclay, 91191 Gif-sur-Yvette, France}
\affiliation{Sorbonne Universit\'e, CNRS, Laboratoire de Physique Th\'eorique de la Mati\`ere Condens\'ee, LPTMC, F-75005 Paris, France}
\email{cesare.nardini@gmail.com }


\maketitle

\section{Further numerical results}
The numerics shown in the main text have all been performed using a pseudo-spectral scheme, using both in-house developed Python code, and cuPSS \cite{caballero2024cupss}. The particular solvers in the cuPSS library are available on request.

Every curve in Fig. 1 and Fig. 2 of the main texts as well as Figs 1 and 2 of this SM are obtained by averaging over between $10^2$ and $10^3$ different noise realizations, depending on the system size. 
In all runs, we have set $\gamma = 1$ and $D=0.01$ while other parameters have been defined in the text. The results have been presented in terms of the non-dimensional coupling constant $\bar\alpha = \alpha D^{1/2}\gamma^{-3/2}$ so that $\alpha = 10 \bar\alpha$. 

\begin{figure}[h!]
    \centering
    \includegraphics[scale=0.45]{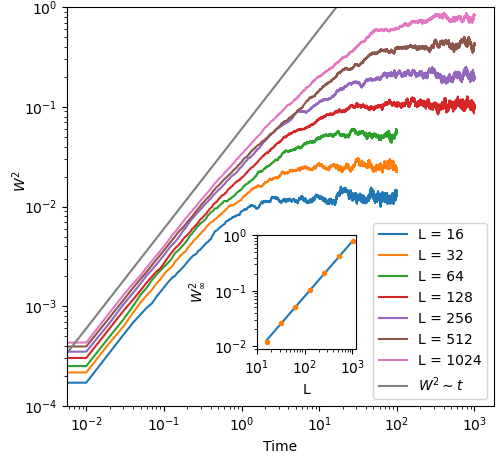}
    \includegraphics[scale=0.48]{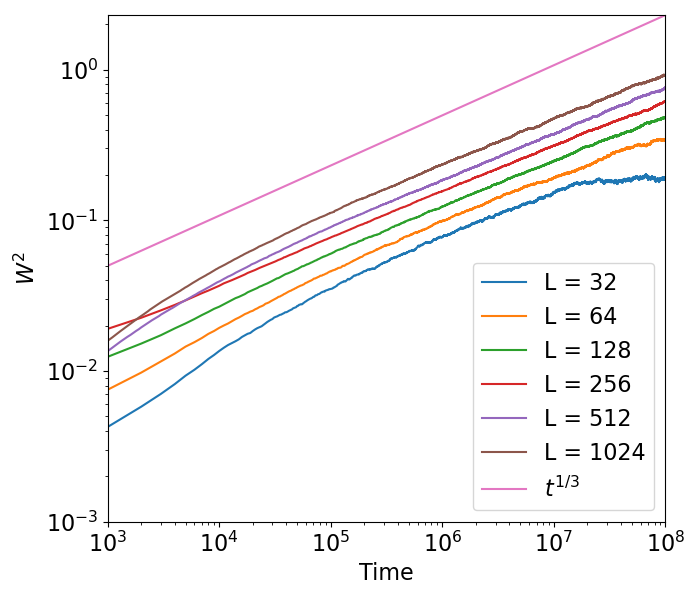}
    \caption{Simulations of eq. (12) of the main text for (Left) $\lambda=0$ and $\alpha=0$ showing the expected results from the linear theory. Simulations of eq. (12) of the main text for (Right) $\lambda=10^3$ and $\alpha=0$.}
    \label{fig:w_linear}
\end{figure}

Fig. \ref{fig:w_linear}(Left) report numerical simulations of Eq. (12) of the main text for $\lambda=0$ and $\alpha = 0$, for which we expect linear growth $W^2$ at early times, and the saturated value $W_\infty^2$ to linearly increase with system size. This is indeed what we observe in Fig. \ref{fig:w_linear}(Left). Fig. \ref{fig:w_linear}(Right) shows numerics of Eq. (12) with $\lambda=10^3$ and $\alpha=0$. As expected, we now observe $W^2\sim t^{1/3}$.

\begin{figure}[h!]
    \centering
    \includegraphics[scale=0.6]{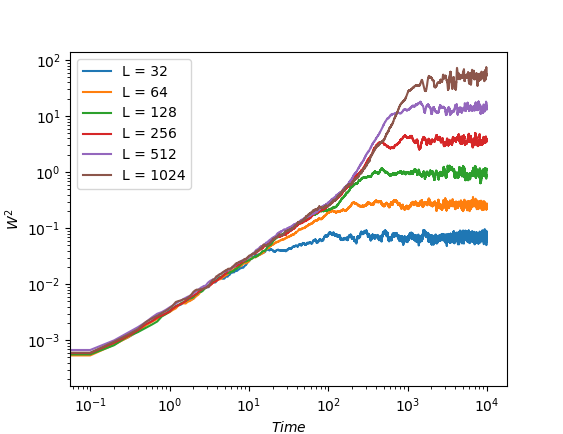}
    \includegraphics[scale=0.6]{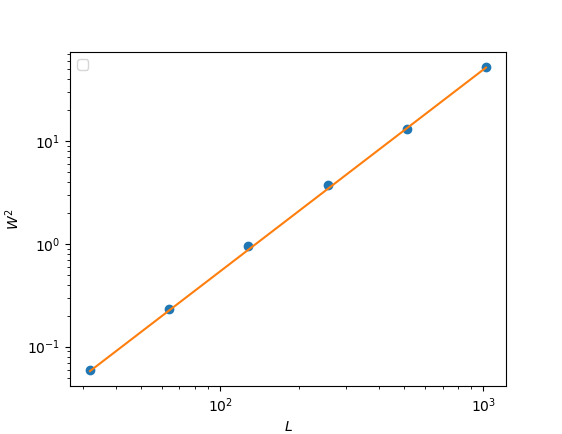}
    \caption{Simulations of Eq. (12) in the main text for $\lambda=10^{-3}$ and various system sizes yield compatible results with those presented in Fig. 2b and 2c of the main text. The slope in the linear fit of the second figure gives $2\chi\approx1.96$, very close to the value of $\chi=1$ that we find in the strict case of $\lambda = 0$.}
    \label{fig:w_Fig2b}
\end{figure}

In Fig. \ref{fig:w_Fig2b} we report the analogues of Figs. 2b and 2c of the main text, now obtained from the integration of Eq. (12) of the main text with $\lambda=10^{-3}$. As shown, the results are fully compatible both for the time evolution and for the scaling of the saturated width. We measured $\chi\simeq 1$.

\section{Derivation of Eq. (12) in the main text for normal active forces}

This section shows the derivation of Eq. (12) of the main text, using the force density $\bff^a$ in Eq. (6) of the main text. Expanding this force density to quadratic order in $\nabla h$, we obtain that $\bff^a = 2\alpha_f[\hat{ y} - \nabla h -\frac{1}{2}\hat{y} (\nabla h)^2]\delta(y-h({\bf x}))$. Notice the first two terms can be written as a pure gradient and, therefore, can be absorbed into a pressure proportional to $\phi$. Thus, they do not lead to a flow in an incompressible system. We are thus left with $\bff^a = -\alpha_f\hat{y}(\nabla h)^2\delta(y-h({\bf x})) +O(h^3)$.
Additionally, there is an equilibrium capillary force density $\bff^p = -\phi\nabla\mu$, that generates a force density responsible for interfacial tension \cite{bray2001interface}.

Writing Eq. (8) of the main text with both the equilibrium and nonequilibrium forces, we obtain
\qq
-\varphi'(u)\partial_th(\bfx,t) = -\varphi'(u)\int d\bfx'dy' T_{yj}(\bfx-\bfx', y-y') \left(f^p_j(\bfx', y') + f^a_j(\bfx',t)\right),
\qqq
where we have defined $u=y-h(\bfx,t)$. We can now substitute the values of the two forces to leading order in $h(\bfx,t)$. Notice that the first term in the force density, which is the equilibrium capillary force, is responsible for the linear interfacial tension, the only contribution that remains in equilibrium. Notice as well that, as explained in the paragraph above, the leading order in $h(\bfx, t)$ contribution to $\bff^a$ is already quadratic in $\nabla h$. With this in mind, we write
\qq
-\varphi'(u)\partial_t h(\bfx,t) = -\gamma\varphi'(u)\int d\bfx' T_{yy}(\bfx-\bfx',0)\nabla^2h(\bfx',t) +\alpha_f\varphi'(u)\int d\bfr'T_{yy}(\bfx-\bfx',y-h(\bfx',t))(\nabla h(\bfx',t))^2 + \mathcal{O}(h^3),
\qqq
Again, the first term is the equilibrium one, while the second is the only non-vanishing contribution to quadratic order in gradients and fields. We now multiply the equation with $\varphi'(u)$ and integrate across the interface to obtain
\qq
\partial_th(\bfx,t) = \gamma\int d\bfx' T_{yy}(\bfx-\bfx',0)\nabla^2h(\bfx',t) -\alpha_f\int d\bfx'dy' T_{yy}(\bfx - \bfx', h(\bfx,t)-h(\bfx',t))[\nabla h(\bfx',t)]^2 + \mathcal{O}(h^3).
\qqq
We then expand $T_{yy}$ in $h$ in the second term of the r.h.s. and keep only the zeroth order contribution, since this already generates the lowest order term quadratic in $h$. We obtain
\qq
\partial_th(\bfx,t) = \gamma\int d\bfx' T_{yy}(\bfx-\bfx',0)\nabla^2h(\bfx',t) -\alpha_f\int d\bfx'dy' T_{yy}(\bfx - \bfx', 0)(\nabla h(\bfx',t))^2.
\qqq
which is Eq. (12) of the main text written in real space. 

\section{Quadratic nonlinearities are not generated by a stress}\label{app:stress}
This section shows that the nonequilibrium stress
\qq\label{eq:non_eq_stress}
\sigma_{ij} = -k\partial_i\phi\partial_j\phi - \kappa(\phi)\partial_i\phi\partial_j\phi,
\qqq
does not generate quadratic nonlinearities for any local function $\kappa(\phi)$, in the sharp-interface limit. Notice that the equilibrium stress is recovered for $\kappa(\phi) = 0$ and Active Model H is recovered when $\kappa(\phi)$ is constant. 

Consider the equation for $h(\bfx,t)$, as derived from the ansatz $\phi = \varphi(y-h(\bfx,t))$; multiplying Eq. (8) of the main text with $\varphi'(u)$ and integrating across the interface, we obtain:
\qa
\partial_th =-\int d\bfx'dy' T_{yy}(\bfx-\bfx', h(\bfx)-y')\partial_k\sigma_{yk} + \nabla h\int d\bfx'dy' T_{xx}(\bfx-\bfx',h(\bfx)-y')\partial_k\sigma_{xk}.
\qqa
The sharp interface limit further allows us to write the stress as located at the interface, and thus we get
\qa
\partial_th = -\int d\bfx' T_{yy}(\bfx-\bfx',h(\bfx)-h(\bfx'))\int_{-\infty}^\infty dy'\partial_k\sigma_{yk} +\nabla h \int d\bfx' T_{xx}(\bfx-\bfx',h(\bfx)-h(\bfx'))\int_{-\infty}^\infty dy' \partial_k\sigma_{xk}.
\qqa
Observe that the integrals over $y'$ contain terms of the form $\partial_k \sigma_{ij}$, and thus are total derivatives; they vanish when $k=y$ because the stress at infinity vanishes. We further expand each $T_{ij}$ in the height fields, keeping only the lowest order term. We are thus left with
\qa
\partial_th = -\int d\bfx' T_{yy}(\bfx-\bfx',0)\int_{-\infty}^\infty dy'\nabla \sigma_{yx} +\nabla h \int d\bfx' T_{xx}(\bfx-\bfx',0)\int_{-\infty}^\infty dy' \nabla\sigma_{xx}.
\qqa

It can be shown that $\sigma_{xx}$ is of order $(\nabla h)^2$, and thus the second term in the r.h.s. does not contain quadratic terms. Expanding the derivative in the first term of the r.h.s., we obtain
\qa\label{eq:SM-stress}
\nabla\sigma_{xy} = [k+\kappa(\varphi(u'))]\varphi'(u')^2\nabla^2h + k\partial_{u'}[\varphi'(u')^2](\nabla h)^2 + \partial_{u'}[f(\varphi(u'))\varphi'(u')^2](\nabla h)^2.
\qqa
The first term in the r.h.s. is diffusion, albeit renormalized by $\kappa$. The second and third terms are total derivatives in $y'$, and thus do not contribute when integrated over $y'$. Hence, the interface equation reduces to the linear equation (1) of the main text, where we have omitted nonlinearities that are higher order than quadratic in $h$. Our result thus shows that when the stress tensor is as in Eq. (\ref{eq:non_eq_stress}), quadratic nonlinearities do not arise in the interface equation, at least at the bare level. 

\section{Derivation of Eq. (12) for active forces in eq. (5) in the main text}

This section shows that a force density of the form of Eq. (5) of the main text produces a term of the same form as the nonlinearity in Eq. (12) of the main text. Since we have just shown that force densities that can be written as a divergence of a stress cannot yield the nonlinear term in Eq. (12), we only need to consider the force densities with coefficients $K_1$ and $K_2$.
We consider the $y$ component of these force densities and expand them using $\phi(\bfx, y, t) = \varphi(y - h(\bfx,t))$, we obtain
\qa
f^a_y & = K_1\phi(\nabla^2\phi)\partial_y\phi + K_2|\nabla\phi|^2\partial_y\phi \\
& =(K_1\varphi(u)\varphi'(u)\varphi''(u) + K_2\varphi'(u)^3)(1+(\nabla h)^2).
\qqa
The terms that do not depend on $h$ account for pressure jumps across the interface, and can thus be ignored. To proceed, we note that $\partial_u(\varphi(u)\varphi'(u)^2) = \varphi'(u)^3 + 2\varphi(u)\varphi'(u)\varphi''(u)$. Using this equality, we can rewrite the previous equation as
\qq
f^a_y = K_2\partial_u(\varphi(u)\varphi'(u)^2)(\nabla h)^2 + (K_1-2K_2)\varphi(u)\varphi'(u)\varphi''(u)(\nabla h)^2.
\qqq
The first term is a total derivative and thus vanishes upon integration across the interface, as shown in the previous section. 
The second term is the same nonlinearity as in Eq. (12) of the main text, when we set $\alpha = (K_1-2K_2)\varphi(u)\varphi'(u)\varphi''(u)$, as discussed in the main text. In the particular case $K_1-2K_2=0$, we can write the force density in Eq. (5) of the main text as the divergence of a stress $\sigma_{ij} \sim \phi(\nabla_i\phi\nabla_j\phi-\delta_{ij}(\nabla\phi)^2/2)$, that is equivalent to including higher-order terms in the stiffness $\kappa$ in the free energy, i.e. $\kappa \rightarrow \kappa + \kappa'\phi$. This is compatible with the result of Appendix \ref{app:stress}.

\section{Renormalization diagram calculations}

We show that starting with Eq. (17) of the main text, we generate new noise with variance as in Eq. (18) in the main text via a $1$-loop renormalisation calaculation. We use the following notation for the vertex in Eq. (17) of the main text:
\begin{figure}
    \centering
    \includegraphics[scale=0.7]{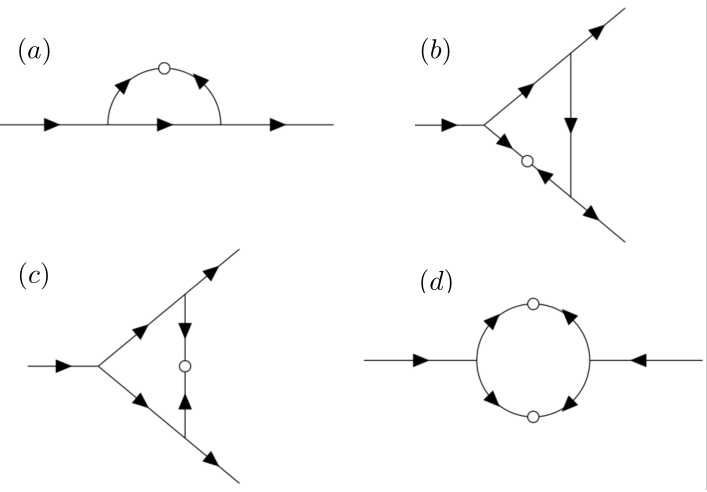}
    \caption{One loop Feynman diagrams for an equation with quadratic nonlinearities.}
    \label{fig:feynamn-diags}
\end{figure}
\qq
v(\bfq_1,\bfq_2,\bfq_3) = \frac{\alpha}{4\eta|\bfq_1|}\bfq_2\cdot\bfq_3,
\qqq
the following ones for the linear propagator $G_0(\bfq,\omega)$ and correlator $C_0(\bfq,\omega)$:
\qa
\label{Eq:G0CO}
G_0(\bfq,\omega) =  \frac{1}{-i\omega+\gamma|\bfq|/4\eta};\qquad
C_0(\bfq,\omega) =  \frac{2D}{4\eta|\bfq|(\omega^2+\gamma^2\bfq^2/16\eta^2)}.
\qqa




With this, we evaluate the Feynman diagram that renormalizes the noise, depicted in Fig \ref{fig:feynamn-diags}d. Calling this diagram (which has a combinatorial factor of $2$) $D_n$, we get
\qq
D_n = 2\int_{-\infty}^\infty\frac{d\omega_I}{2\pi}\int_{|\bfq_I|\in (\Lambda/b,\Lambda)} \frac{d^d\bfq_I}{(2\pi)^{d}} C_0(\bfq_I,\omega_I)C_0(\bfq-\bfq_I,-\omega_I)v(\bfq,\bfq_I,\bfq-\bfq_I)v(-\bfq,-\bfq_I,-\bfq+\bfq_I)\,.
\qqq
Expanding the integrand to the lowest order in $|\bfq|$ and performing the integrals, we obtain
\qq\label{eq:app:Dn}
D_n = \frac{2D\bar\alpha^2\Lambda^{d-1}}{4\eta|\bfq|^2}dbS_d + O(|\bfq|^{-1})\equiv\frac{2D'}{|\bfq|^2}+ O(|\bfq|^{-1}).
\qqq
where $S_d = \Omega_d/(2\pi)^d$, with $\Omega_d$ being the surface of a $d$-dimensional sphere, and $db$ is an infinitesimally thin shell of modes that we are integrating out, $b\rightarrow 1+db$. $D'$ is the coefficient introduced in Eq. (18) of the main text.
Thus, Eq. \eqref{eq:app:Dn} shows that a noise with variance proportional to $\bfq^{-2}$ is generated, as stated in the main text.

Note that if we evaluate the noise graph using the linear correlator and vertex function of Eq. (12) of the main text, the leading order wavenumber-dependence is given by $\mathcal{T}({\bf q},0)^2$; therefore, in $\lambda\gg|\bfq|$ limit, it vanishes as $|\bfq|^2$ at small $|\bfq|$ and is subleading to the bare noise in this limit. Conversely, it generates a noise whose correlation is $\sim 1/|\bfq|^2$ at small $|\bfq|$ in the $\lambda\to 0$ limit, consistent with the calculation presented here.

\section{$1$-loop analysis of Eq. (12) in the main text and singularity of the $\lambda\to0$ limit}
In this section, we evaluate i. the propagator diagram for Eq. (12) of the main text; ii. the propagator diagram for Eq. (17) of the main text and iii. the propagator diagram from Eq. (17) of the main text when the noise correlation is given by Eq. (18) of the main text. 

To evaluate i., we need the linear propagator and correlator, and the vertex for Eq. (12) of the main text which are
\qa
G^F_0(\bfq,\omega) =  \frac{1}{-i\omega+\gamma\,\mathcal{T}({\bf q},0)|{\bf q}|^2}\qquad;\qquad
C^F_0(\bfq,\omega) =  \frac{2D\, \mathcal{T}({\bf q},0)}{\omega^2+\gamma^2\mathcal{T}({\bf q},0)^2|\bfq|^4};\qquad v^F(\bfq_1,\bfq_2,\bfq_3) = {\alpha}\mathcal{T}({\bf q},0)\,\bfq_2\cdot\bfq_3\,.
\qqa
Defining
\begin{equation}
    {\bf q}_\pm={\bf q}_I\pm\frac{{\bf q}}{2},
\end{equation}
as is standard in calculating the propagator renormalisation in the KPZ equation \cite{kardar1986dynamic}, we obtain the expression for the diagram in Fig. \ref{fig:feynamn-diags} a (which has a combinatorial factor 4):
\begin{equation}
\label{eq:intprop}
I=4\int_{-\infty}^\infty\frac{d\omega_I}{2\pi}\int_{|\bfq_I|\in (\Lambda/b,\Lambda)} \frac{d^d\bfq_I}{(2\pi)^{d}} C^F_0(\bfq_-,\omega_I)G^F_0(\bfq_+,\omega-\omega_I)v^F(\bfq,\bfq_+,-\bfq_-)v^F(\bfq_+,\bfq_-,\bfq)\,.
\end{equation}
To evaluate this integral, we need to expand in $|\bfq|$; when we do that at a finite $\lambda$, we are implicitly in the $|\bfq|\ll\lambda$ limit.  In this limit,
\begin{equation}
    I=\mathcal{T}({\bf q},0)\left[\frac{(d-3)\bar{\alpha}^2\gamma}{d}\Lambda^{d-2}S_d|\bfq|^2db+\mathcal{O}(|\bfq|^3)\right]=\frac{(d-3)\bar{\alpha}^2\gamma}{2\eta\lambda^2d}\Lambda^{d-2}S_d|\bfq|^3db+\mathcal{O}(|\bfq|^4)\,,
\end{equation}
 where, in the second equality, we have used the expression of $\mathcal{T}({\bf q},0)$ in the $|\bfq|\ll\lambda$ limit. Note that we obtain the same leading order correction if we start from the $|\bfq|$KPZ equation displayed in Eq. (15) of the main text. In the $|\bfq|$KPZ equation, this implies a correction to $\gamma$: $\gamma\to \gamma[1+(3-d)\bar{\alpha}^2\Lambda^{d-2}S_d/2\eta\lambda^2d]$.

 We can also evaluate the integral in \eqref{eq:intprop} by first taking the $\lambda\to 0$ limit in the integrand and then the $|\bfq|\to 0$ limit. In this case, we obtain
 \begin{equation}
    I=\mathcal{T}({\bf q},0)\left[\frac{(d-1)\bar{\alpha}^2\gamma}{d}\Lambda^{d-2}S_d|\bfq|^2db+\mathcal{O}(|\bfq|^3)\right]=\frac{(d-1)\bar{\alpha}^2\gamma}{4\eta d}\Lambda^{d-2}S_d|\bfq|db+\mathcal{O}(|\bfq|^2)\,,
\end{equation}
 where, in the second equality, we have used the expression of $\mathcal{T}({\bf q},0)$ in the $\lambda\to 0$ limit. We obtain the same leading order correction if we start from Eq. (17) of the main text; thus, we have evaluated ii. above. Note that this is again a correction to $\gamma$ in Eq. (17); no lower order term in $|\bfq|$ is generated. 
 
Expanding $\mathcal{T}({\bf q},0)$ in the small $\lambda$ limit and going beyond leading (zeroth) order in $\lambda$, we get $1/4\eta|{\bf q}|-3\lambda^2/16\eta|{\bf q}|^3+\mathcal{O}(\lambda^4)$. Therefore, if starting from a bare theory with $\lambda=0$, linear terms $\propto h_{\bfq}/|\bfq|^{2n-1}$ with $n>0$ emerged from the propagator diagram, we could argue that an effective $\lambda$ was being generated. No such term emerges. However, the expansion of $\mathcal{T}({\bf q},0)$ for small $\lambda$ also clearly demonstrates that $\lambda$, if present, is relevant. Therefore, the $\lambda=0$ limit is singular: a model with arbitrarily small---but non-zero---$\lambda$ flows away from $\lambda=0$ fixed point.

Finally, we evaluate the propagator diagram for Eq. (17) of the main text but with the noise correlator being given by Eq. (18) of the main text (our item iii. above). The propagator (same as in Eq. \eqref{Eq:G0CO} of the last section), and the correlator (which is now different) are given by
\qa
G_0(\bfq,\omega) =  \frac{1}{-i\omega+\gamma|\bfq|/4\eta};\qquad
C^N_0(\bfq,\omega) =  \frac{2D'}{|\bfq|^2(\omega^2+\gamma^2\bfq^2/16\eta^2)}\,.
\qqa
The integral is
\begin{multline}
I_2=4\int_{-\infty}^\infty\frac{d\omega_I}{2\pi}\int_{|\bfq_I|\in (\Lambda/b,\Lambda)} \frac{d^d\bfq_I}{(2\pi)^{d}} C^N_0(\bfq_-,\omega_I)G_0(\bfq_+,\omega-\omega_I)v(\bfq,\bfq_+,-\bfq_-)v(\bfq_+,\bfq_-,\bfq)\\=\frac{(d-2)\bar{\alpha}^2D'\gamma}{D\eta d}\Lambda^{d-3}S_d|\bfq|db+\mathcal{O}(|\bfq|^2)\,.
\end{multline}
This still scales as $\sim |\bfq|$ demonstrating that even the more strongly divergent noise correlator doesn't yield a lower order in $|\bfq|$ linear term.


\bibliography{apssamp}